\documentclass[acmsmall,dvipsnames,table]{acmart}

\AtBeginDocument{}

\setcopyright{acmcopyright}
\copyrightyear{2018}
\acmYear{2018}
\acmDOI{XXXXXXX.XXXXXXX}

\acmJournal{TOSEM}
\acmPrice{15.00}
\acmISBN{978-1-4503-XXXX-X/18/06}

\usepackage{algorithmic}
\usepackage{graphicx}
\usepackage{textcomp}
\usepackage{url}
\usepackage{cleveref}
\usepackage{colortbl}
\usepackage{booktabs} 
\usepackage{caption}
\usepackage[nolist,nohyperlinks]{acronym}
\usepackage{subcaption}
\usepackage{multirow}
\usepackage{xspace}
\usepackage{sphack}
\usepackage{enumitem} 
\usepackage{soul}

\begin{acronym}
\acrodef{C}{Coverage}
\acrodef{HV}{Hypervolume}
\acrodef{IGD+}{Inverted Generational Distance Plus}
\acrodef{EP}{Epsilon}
\acrodef{pas}{performance antipattern}
\end{acronym}

\newcommand{\pas}{\textit{\ac{pas}}\xspace}
\newcommand{\easier}{\textit{Easier}\xspace}
\newcommand{\perfq}{\textit{performance}\xspace}
\newcommand{\achanges}{\textit{cost}\xspace}
\newcommand{\reliability}{\textit{reliability}\xspace}
\newcommand{\ie}{\emph{i.e.,}\xspace}
\newcommand{\eg}{\emph{e.g.,}\xspace}

\newcommand{\ttbs}{\texttt{TTBS}\xspace}
\newcommand{\ccm}{\texttt{CoCoME}\xspace}

\newcommand{\brf}{\textit{BRF}\xspace}

\newcommand{\expComp}[1]{\textbf{#1} UML Components\xspace}
\newcommand{\expNode}[1]{\textbf{#1} UML Nodes\xspace}
\newcommand{\expUC}[1]{\textbf{#1} UML Use Cases\xspace}
\newcommand{\nsga}{\textit{NSGA-II}\xspace}
\newcommand{\hv}{$HV$\xspace}
\newcommand{\igd}{$IGD\text{+}$\xspace}
\newcommand{\eps}{$Epsilon$\xspace}

\newcommand{\refpf}{$Reference_{1000}$\xspace}
\newcommand{\base}{$Baseline_{100}$\xspace}

\newcommand{\ttsecondintcentrA}{$Interactive^{2nd (c258)}_{50}$\xspace}
\newcommand{\ttsecondintcentrB}{$Interactive^{2nd (c223)}_{50}$\xspace}

\newcommand{\ccsecondintcentrA}{$Interactive^{2nd (c317)}_{50}$\xspace}
\newcommand{\ccsecondintcentrB}{$Interactive^{2nd (c358)}_{50}$\xspace}

\usepackage[most]{tcolorbox}
\newtcolorbox{rqsummary}[1]{
  enhanced,
  top=1em,
  fonttitle=\bfseries,
  coltitle=black,
  colback=white,
  attach boxed title to top left={xshift=1em,yshift=-\tcboxedtitleheight/2},
  boxed title style={size=small,colback=white},
  title={#1},
}

\usepackage{ifthen}
\newboolean{showcomments}
\setboolean{showcomments}{false}

\ifthenelse{\boolean{showcomments}}
  {
	
	  \newcommand{\nb}[2]{
    		\fbox{\bfseries\sffamily\scriptsize#1}
		{\sf\small$\blacktriangleright$\textit{#2}$\blacktriangleleft$}
   	  }
  }
  {
	\newcommand{\nb}[2]{}

}

\newcommand\DDP[1]{\textcolor{SeaGreen}{\nb{Daniele}{#1}}}

\newcommand\ADP[1]{\textcolor{RawSienna}{\nb{Andres}{#1}}}

\newcommand\PJ[1]{\textcolor{Turquoise}{\nb{Pooyan}{#1}}}

\let\oldciteauthor=\citeauthor
\renewcommand\citeauthor[1]{\protect\NoHyper\oldciteauthor{#1}\protect\NoHyper} 

\newcommand{\SoBigDataITAck}{European Union --- NextGenerationEU --- National Recovery and Resilience Plan (Piano Nazionale di Ripresa e Resilienza, PNRR) --- Project: ``SoBigData.it --- Strengthening the Italian RI for Social Mining and Big Data Analytics'' --- Prot. IR0000013 --- Avviso n. 3264 del 28/12/2021.\xspace}

\newcommand{\dqualizer}{German Federal Ministry of Education and Research (BMBF), project: ``dqualizer: Domain-centric Runtime Quality Analysis of Business-critical Application Systems'', grant number 01IS22007A/B.\xspace}

\begin{document}

\title{Introducing Interactions in Multi-Objective Optimization of Software Architectures}

\author{Vittorio Cortellessa}
\email{vittorio.cortellessa@univaq.it}
\orcid{0000-0002-4507-464X}
\affiliation{\institution{University of L'Aquila}
  \city{L'Aquila}
  \country{Italy}
}

\author{J. Andres Diaz-Pace}
\email{andres.diazpace@isistan.unicen.edu.ar}
\orcid{0000-0002-1765-7872}
\affiliation{\institution{ISISTAN, CONICET-UNICEN}
  \city{Tandil}
  \country{Argentina}
}

\author{Daniele {Di Pompeo}}
\orcid{0000-0003-2041-7375}
\email{daniele.dipompeo@univaq.it}
\affiliation{\institution{University of L'Aquila}
  \city{L'Aquila}
  \country{Italy}
}

\author{Sebastian Frank}
\email{sebastian.frank@uni-hamburg.de}
\orcid{0000-0002-3068-1172}
\affiliation{\institution{University of Hamburg}
  \city{Hamburg}
  \country{Germany}
 }
 
\author{Pooyan Jamshidi}
\email{pjamshid@cse.sc.edu}
\orcid{0000-0002-9342-0703}
\affiliation{\institution{University of South Carolina}
  \country{USA}
 }

\author{Michele Tucci}
\email{michele.tucci@univaq.it}
\orcid{0000-0002-0329-1101}
\affiliation{\institution{University of L'Aquila}
 \city{L'Aquila}
 \country{Italy}}

\author{Andr\'e van Hoorn}
\authornote{\em{This work is dedicated to the memory of our co-author and friend, Andr\'e, whose brilliance, dedication, and kindness left an indelible mark on all of us.}}
\email{andre.van.hoorn@uni-hamburg.de}
\orcid{0000-0003-2567-6077}
\affiliation{\institution{University of Hamburg}
  \city{Hamburg}
  \country{Germany}
 }
 
\renewcommand{\shortauthors}{Cortellessa et al.}

\begin{abstract}

Software architecture optimization aims to enhance non-functional attributes like performance and reliability while meeting functional requirements. 
Multi-objective optimization employs metaheuristic search techniques, such as genetic algorithms, to explore feasible architectural changes and propose alternatives to designers. However, this resource-intensive process may not always align with practical constraints.

This study investigates the impact of designer interactions on multi-objective software architecture optimization. 
Designers can intervene at intermediate points in the fully automated optimization process, making choices that guide exploration towards more desirable solutions. 
Through several controlled experiments as well as an initial user study (14 subjects), we compare this interactive approach with a fully automated optimization process, which serves as a baseline.
The findings demonstrate that designer interactions lead to a more focused solution space, resulting in improved architectural quality. 
By directing the search towards regions of interest, the interaction uncovers architectures that remain unexplored in the fully automated process.
In the user study, participants found that our interactive approach provides a better trade-off between sufficient exploration of the solution space and the required computation time. 
\end{abstract}



\maketitle

\section{Introduction}\label{sec:intro}

Historically, software architecture refactoring was a manual, labor-intensive process. Designers and architects would carefully analyze the architecture to identify bottlenecks or inefficiencies and then manually apply refactorings to improve non-functional properties such as performance, reliability, and maintainability~\cite{DBLP:journals/sqj/BaqaisA20}. This manual approach, though effective in leveraging human expertise, was time-consuming, prone to error, and difficult to scale, particularly in large, complex software systems~\cite{DBLP:journals/tse/KimZN14}.

With advances in machine learning and optimization algorithms, the software engineering community began shifting toward fully automated processes to tackle refactoring challenges~\cite{DBLP:journals/sqj/BaqaisA20}. Automated optimization techniques, such as evolutionary algorithms, allow for systematic exploration of the design space, offering scalable and efficient ways to optimize software architectures~\cite{DBLP:journals/ese/RamirezRV16}. These methods generate and evaluate multiple solutions in parallel, freeing designers from the repetitive and painstaking task of manual refactoring. As a result, automated refactoring approaches have gained significant traction, enabling faster iterations and more consistent results across different types of systems.

Although fully automated approaches brought about efficiency, they introduced new limitations. The rigid rule-based nature of these algorithms often did not account for the nuanced domain-specific knowledge that human architects possess~\cite{alotaibi2018advances}. This knowledge is especially critical when trade-offs between conflicting objectives, such as performance and maintainability, must be balanced. Therefore, while the shift towards full automation represented a significant advancement, it also highlighted the need to reintroduce human judgment into the loop~\cite{ramirez2018interactive,kharchenko2014method}.

This paper proposes a hybrid approach that combines the best of both worlds, leveraging the computational power of optimization algorithms while incorporating human interaction to guide the process. By allowing designers to interact with intermediate solutions, we aim to capture the evolving preferences and contextual insights that fully automated methods miss. This integration addresses the broader need for practical decision-making in software architecture, where both human expertise and automated efficiency play crucial roles.

Multi-objective black-box optimization addresses the simultaneous optimization of several objectives, through a black-box interface, through the search for Pareto optimal solutions within a vast search space ~\cite{ii2019practical}. This optimization category presumes that the optimization function is neither analytically accessible nor smooth, thus preventing the use of common gradient-based optimization techniques such as the Newton-Raphson method ~\cite{boyd2004convex}. Essentially, the optimization algorithm is limited to querying the function value $f(x)$ at a specific point $x$ through the black-box interface.
Therefore, it is necessary to employ design space exploration techniques, which are typically implemented as fully automated iterative processes without the provision for human interaction. These processes are set up with certain parameters (\eg number of iterations) and, in evolutionary methods, include mechanisms for the automatic combination of intermediate solutions (\ie crossover and mutation operators). A wide range of issues in software engineering~\citep{Ramirez:2018uz,Mariani:2017jd,Ouni:2017db,Ray:2014ip,Bavota:2014kr,Kessentini:2012cb,chen2021efficient}, computer systems~\cite{alipourfard2017cherrypick,li2020statically,JC:MASCOTS16,iqbal2022unicorn}, and hyperparameter tuning \citep{snoek2012practical,shahriari2015taking} have been effectively tackled using multi-objective optimization approaches. These issues typically involve multiple metrics to measure non-functional attributes being targeted as optimization objectives~\citep{Aleti:2013gp,Martens:2010bn,DBLP:conf/mompes/AletiBGM09,DBLP:journals/csur/BlumR03}.

Evolutionary algorithms have often been used for tackling multi-objective optimization problems in the software architecture refactoring domain. These algorithms iteratively generate and combine architectural models with specified rules to explore the solution space, while taking into account user-defined constraints that prevent unfeasible solutions from being considered. However, this process cannot guarantee that the solutions obtained will be satisfactory to the designer. The key reason is that the designer may have some preferences that are not well understood at the beginning of the process, but they may be clarified by glimpsing at intermediate solutions.
In fact, the introduction of designer interactions in an optimization process has been shown to help drive it towards more desirable solutions~\citep{aljawawdehMetaheuristicDesignPattern2015}.
This is because expert domain knowledge, which can be hard to express as initial constraints for the optimization process, can be later conveyed in the process upon direct observation of intermediate solutions.
However, it is important to note that the targeted exploration of the solution space based on the designer’s emerging preferences may lead to solutions that are quantitatively worse in terms of traditional optimization metrics if we are treating all objectives as equally important. Despite this, such solutions may better align with the designer's qualitative priorities or context-specific trade-offs, thus emphasizing the value of human judgment in achieving practical and desirable outcomes.
For instance, the designer might observe that an intermediate solution induces an interesting trade-off among the objectives, and then she can decide to restart the process from that solution in order to focus the search on the solution proximity.

This paper focuses on software architecture refactoring aimed at optimizing non-functional attributes or properties, like performance and reliability. In this context, we pose the central hypothesis that \emph{the interaction of designers with an automated optimization engine can drive the space exploration towards candidate architectures which more likely induce desirable trade-offs among such properties.} 
In other words, this paper aims to show whether enabling designers to interact with an architecture optimization process opens new possibilities for them to improve the process and the quality of solutions. Specifically, we performed several experiments in this context, where an evolutionary algorithm uses refactoring actions to automatically explore the space of software architectural models, and the designer can specify (in intermediate iterations) regions of interest where the exploration should focus on. A region of interest refers to a group of related solutions where a representative trade-off among the optimization objectives can be explored. This scenario contrasts with the traditional approach of a fully automated optimization. 

Different forms of interaction have gained attention in addressing some of the challenges in Search-Based Software Engineering (SBSE) studies, as highlighted by \citet{Ramirez_Romero_Simons_2019}, where several interactive modes have been identified.
Here, we explored only the \textit{preference-based interactivity}, which allows the designer to steer the search toward solutions inducing specific trade-offs, such as low response time and high reliability.
The exploration of other types of interactions is left to future work.

\begin{figure}[htbp]
    \centering
\includegraphics[width=\linewidth]{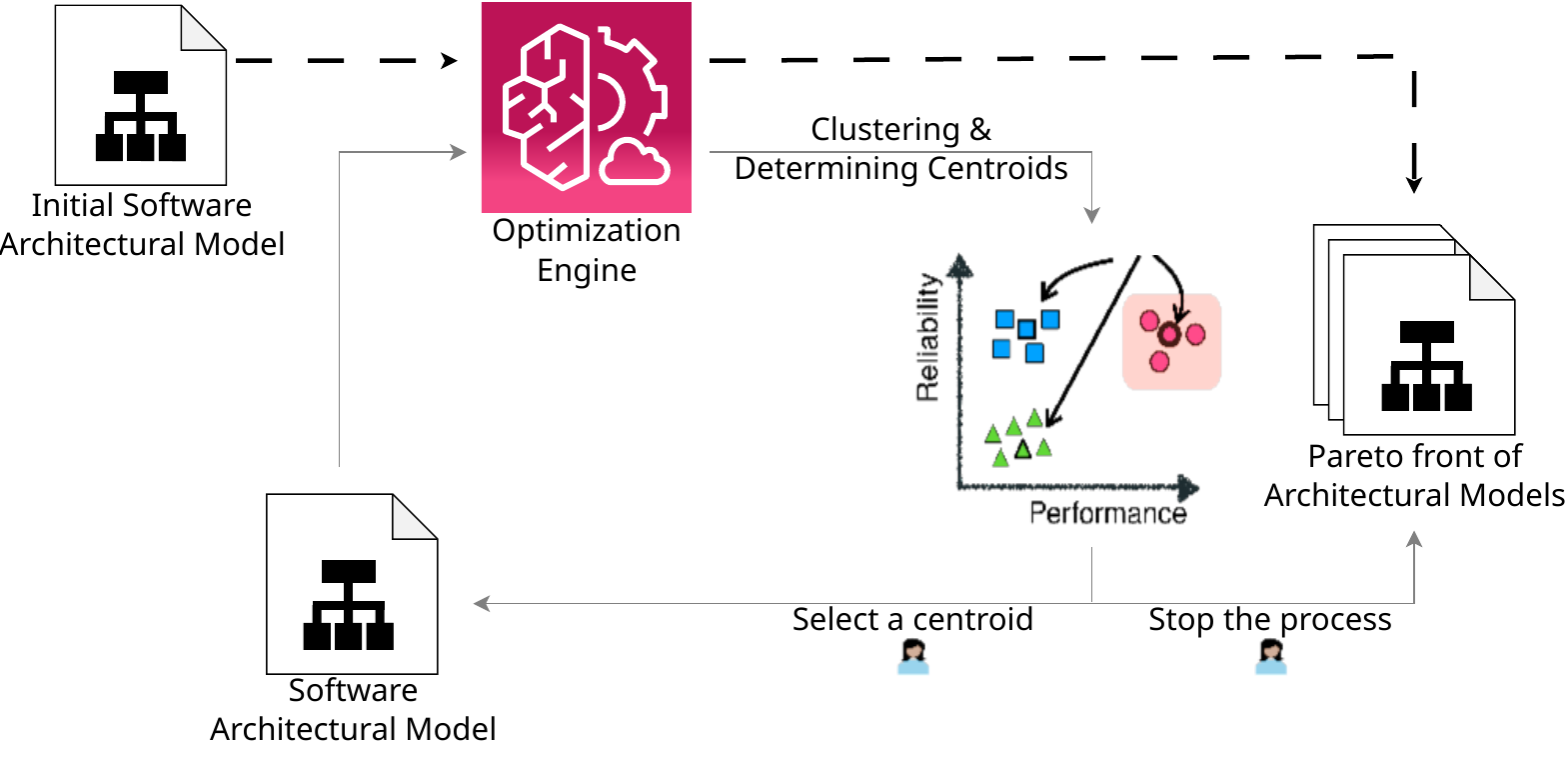}
\caption{A schematic representation of an interactive optimization process. Humans can interact with the process by (1) selecting a centroid of the intermediate clustered Pareto front or (2) stopping the process. }\label{fig:schema}
\end{figure}

A schematic representation of an interactive optimization process is illustrated in \Cref{fig:schema}. The dashed black arrows in the figure represent the flow of a traditional fully automated process, like EASIER~\cite{Arcelli:2018vo}, Nautilus~\cite{ferreira2020nautilus}, MORE~\cite{Ouni_Kessentini_Cinneide_Sahraoui_Deb_Inoue_2017}, PerOpteryx~\cite{Koziolek_Koziolek_Reussner_2011}, ArcheOpterix~\cite{DBLP:conf/mompes/AletiBGM09}, or SQuAT~\cite{RagoSBCARS17}, among others. The gray filled arrows introduce interactivity in the process.

The process starts with an initial software architectural model that feeds the optimization engine. 
In a fully automated process, as search-based approaches generally are, the engine produces a final Pareto front of architectural models after several iterations. 
In this context, we envisage interactions where humans select centroids of automatically generated clusters of solutions within the intermediate Pareto front. The loop stemming from the Optimization Engine in \Cref{fig:schema} represents such interactions that iterate until the human designer decides to stop the interactive process at her convenience.

Several (functional and non-functional) properties (e.g., security and maintainability) of software can be considered as target of an optimization process.
For example, \citet{DBLP:journals/smr/OuniKCSDI17} improved the maintainability of software through refactoring optimization. 
\citet{DBLP:journals/ase/BoukharataOKBW19} optimized the modularity of web service interfaces.
Nevertheless, in this paper, the optimization objectives refer to: (i) performance and reliability as non-functional properties, and (ii) cost of refactoring and number of performance antipatterns as architectural properties, since they have demonstrated a comprehensive but conflicting combination of quality attributes~\cite{DBLP:conf/icsa/BuschFK19,DBLP:journals/tse/AletiBGKM13,CORTELLESSA2023107159}.

In order to evaluate our central hypothesis, we designed controlled experiments and a preliminary user study (14 subjects) to answer the following research questions:
\begin{itemize}
\item $RQ1$: \textit{To what extent can an interactive process affect the quality of solutions?}
\item $RQ2$: \textit{How different are, in terms of architectural properties, the solutions generated through an interactive process with respect to the ones generated through a fully automated one?}
\item $RQ3$: \textit{How does an interactive process impact the coverage of the solution space?} 
\item $RQ4$: \textit{How do human software architects perceive the interactive process with regard to its usefulness of interactive features and benefits compared to a full solution space exploration?}
\end{itemize}

$RQ1$ studies traditional multi-objective quality indicators that quantify the performance of the searching algorithm, whereas $RQ2$ studies how solutions differ in terms of architectural properties, which in our case are represented by cost and number of performance antipatterns. $RQ3$ investigates the coverage of the solution space in order to understand how the interactive process affects its exploration. 
The first three research questions focus on the objectively measurable effects of an interactive process. In $RQ4$, our aim is to consider the human perspective on the
interactive process, investigating whether the characteristics of our interactive process are considered useful and whether participants see the benefits of the interactive approach compared to a fully automated approach.

In our experimental setup, we rely on specific modeling languages for the definition and evaluation of architectural models. Specifically, we use some UML diagrams to represent architectural models, and stochastic notations (\eg Layered Queueing Networks (LQNs) for performance analysis) to evaluate their non-functional properties. Furthermore, we performed a user study with $14$ architects to assess how our approach supports them in exploring alternative solutions in a large design space.

Our experimentation provides evidence that designer's interactions narrow down the explored solution space while leading to more desirable architectures in terms of quality properties.
Moreover, by focusing on regions of interest for the designer, the interactions steer the search toward architectures that were not reached by the fully automated process. The findings of the user study corroborated this aspect and also exposed that designers considered our interactive features useful and appreciated the reduction in computation time compared to a full exploration of the solution space.

This paper is structured as follows: \Cref{sec:background} introduces the key concepts employed in the study, setting the foundation for the research; \Cref{sec:related-work} reports related studies; \Cref{sec:design} outlines the study design, emphasizing the interactive process and presenting the research questions that guide our investigation; \Cref{sec:results} comprehensively reports the study results, addressing the research questions and shedding light on the outcomes; \Cref{sec:results-userstudy} describes the outcome of the user study; in \Cref{sec:threats}, we examine threats to validity; finally, in \Cref{sec:conclusion}, we conclude the paper by summarizing the study outcomes and providing an overview of future research directions.

 \section{Background}\label{sec:background}

\newcommand{\shorten}[1]{\ignorespaces}

This section introduces the required background for this paper, covering 
 model-based prediction of quality attributes, search-based optimization of quality attributes, the \textit{Easier} evolutionary model-based optimization approach. 

\subsection{Model-based prediction of quality attributes}\label{sec:bg:mb}

As software systems get complex, predicting their quality becomes challenging. For large software systems, it is often too late to assess their quality after the system has already been implemented, because problems are more difficult to fix. 
Therefore, quality prediction based on a system's architectural model is a valuable approach to avoid costs and effort caused by a ``fix-it-later approach''.  
Several approaches have been introduced in the last decades to ease the model-based prediction of quality attributes \citep{DBLP:books/daglib/0027475, reussner2016modeling}.

\begin{figure}
	\centering
\begin{subfigure}[b]{0.7\linewidth}
        	 \centering
        	 \includegraphics[width=\textwidth]{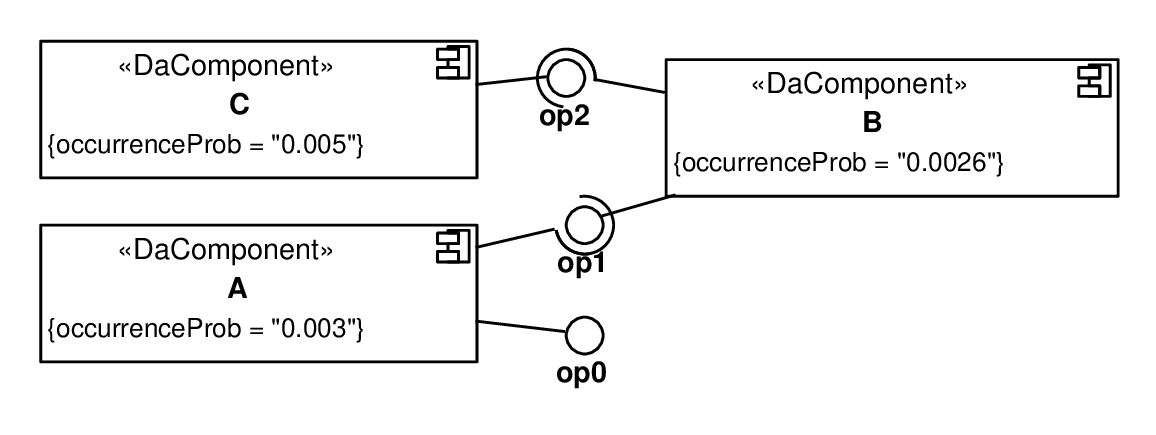}
        	 \caption{Component diagram}
        	 \label{fig:component}
     	\end{subfigure}
     	\hfill
\begin{subfigure}[b]{0.7\linewidth}
        	 \centering
        	 \includegraphics[width=\textwidth]{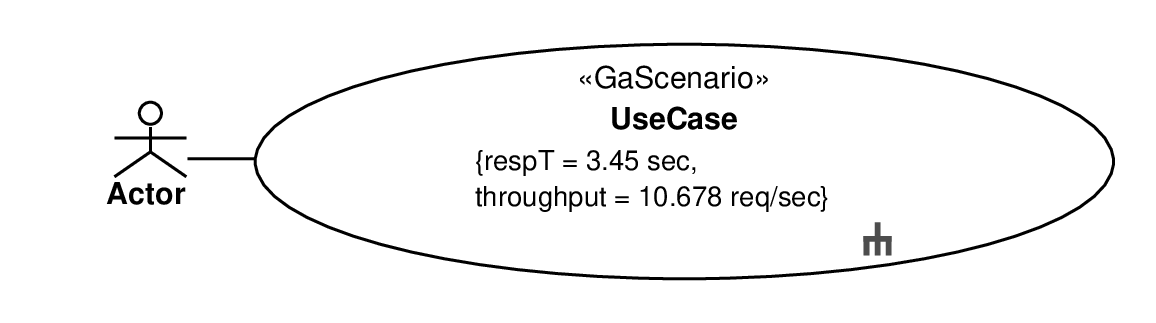}
        	 \caption{Use Case diagram}
        	 \label{fig:usecase}
     	\end{subfigure}
     	\hfill
     	\begin{subfigure}[b]{.9\linewidth}
        	 \centering
        	 \includegraphics[width=\textwidth]{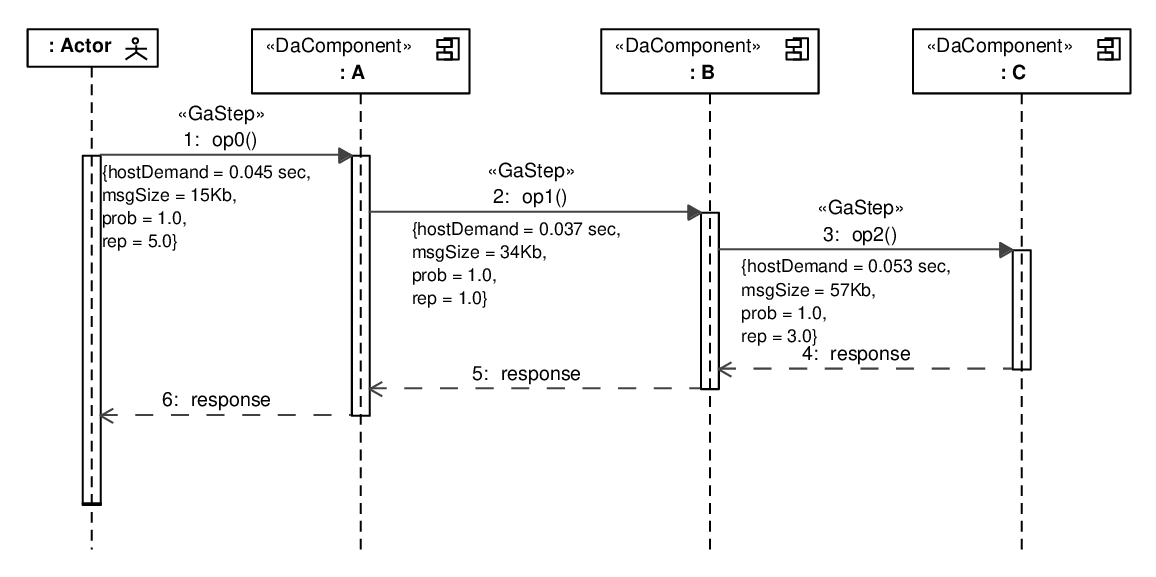}
        	 \caption{Sequence diagram}
        	 \label{fig:sequence}
     	\end{subfigure}
        \hfill 
    	\begin{subfigure}[b]{0.7\linewidth}
        	 \centering
        	 \includegraphics[width=\textwidth]{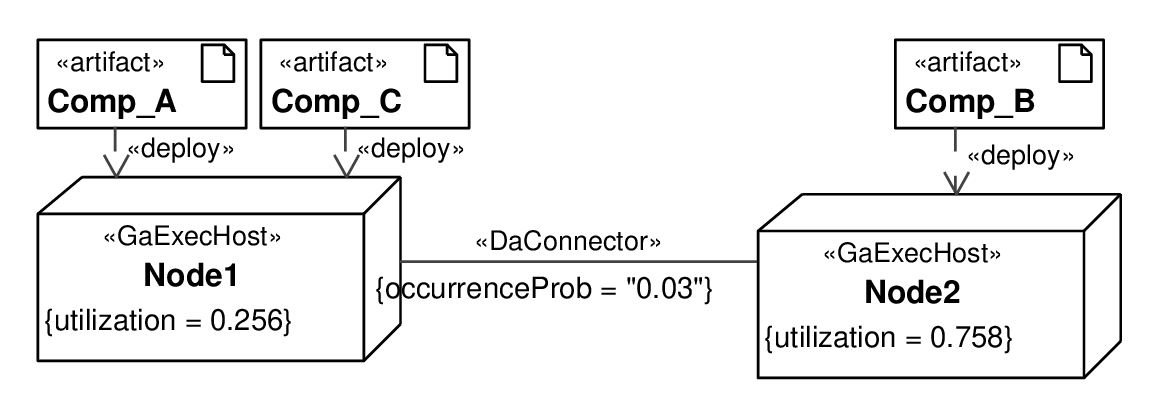}
        	 \caption{Deployment diagram}
        	 \label{fig:deployment}
     	\end{subfigure}
	\caption{Profiled UML model example.}
	\label{fig:uml-example}
\end{figure}

This paper is particularly interested in approaches that allow the modeling of quality attributes at an architectural level using design models. 
We focus on performance and reliability modeling using UML and respective profiles, namely MARTE~\cite{MARTE} and DAM~\cite{BernardiMP11}. 
\Cref{fig:uml-example} shows an example of using UML with the two profiles.
The Use Case diagram (\Cref{fig:usecase}) defines a user scenario, which is detailed by a Sequence diagram (\Cref{fig:sequence}) that captures message exchange among lifelines.
Lifelines represent the components involvement in user scenarios. 
The Deployment diagram (\Cref{fig:deployment}) models platform information and maps components to nodes through artifact manifestations. 
The \emph{DAM.DaConnector}, and \emph{DAM.DaComponent} stereotypes with their tagged values model inputs for the subsequent non-functional analysis, while the \emph{MARTE.GaStep}, \emph{MARTE.GaScenario}, and \emph{MARTE.\-GaExecHost} stereotypes report non-functional analysis results.
A complementing Component diagram (\Cref{fig:component}) allows designing and annotating static connections among components by means of interface realizations and their usages. 

Often, a model-to-model (M2M) transformation is necessary to transform an architectural model into a quantitative one to assess its quality.
For instance, Layered Queueing Networks (LQNs) are a common stochastic modeling formalism that allows predicting the performance of architectural models~\citep{Franks_Al-Omari_Woodside_Das_Derisavi_2009}.
Stochastic models, such as Petri nets and Markov chains, can be employed to quantify the reliability of architectural models~\citep{Cortellessa_Eramo_Tucci_2020}. 

Therefore, in our proposed interactive approach, we exploit: (i) software architectural models expressed in UML, conforming to the four diagrams mentioned above, (ii) LQN to extract performance metrics, and (iii) a closed model to calculate reliability, as presented by \mbox{\citet{CortellessaSC02}}.

\subsection{Search-based optimization of quality attributes}\label{sec:bg:sbse}
Several approaches and tools~\cite{Aleti:2013gp} have been proposed to aid architects in the search-based quality optimization process, \eg PerOpteryx~\cite{Martens:2010bn}, Arche\-Opterix~\cite{DBLP:conf/mompes/AletiBGM09},  SQuAT~\cite{RagoSBCARS17}, and \easier~\cite{Arcelli:2018vo}.
All these approaches require (at least) two inputs: (i)~an initial software architecture model 
and (ii)~objectives that direct the optimization process.
Furthermore, the approaches are equipped with architectural transformations that they can apply to improve the system's quality attributes. 
Such architectural transformations comprise \textit{tactics}~\cite{SoftwareArchitectureinPractice} or \textit{refactorings}~\cite{Arcelli:2018vo}.
A tactic is a domain-specific rule aimed at improving a specific quality attribute, \eg replicating bottleneck services to improve performance~\cite{SoftwareArchitectureinPractice}.
A refactoring, instead, is an architectural transformation that changes the structure of the initial architecture while preserving its behavior, \eg move an internal operation between two components to balance the load of the system~\citep{CORTELLESSA2023107159}.
Multi-objective optimization techniques search the \textit{solution space} to identify the (near-)optimal region of non-dominated solutions, \ie the Pareto front, in terms of objectives quantified by quality metrics~\cite{Aleti:2013gp}.
We consider two classes of objectives: objectives concerning the system's quality, \eg \perfq or \reliability, and objectives referring to the architecture, \eg the number of \textit{performance antipatterns}.
The configuration of the search strategy can determine the shape of the explored space, \ie the diversity, density, and number of solutions. For example, a higher fraction of mutation operations in an evolutionary approach favors more diverse solutions. Furthermore, configurations can also influence the speed and direction of the exploration of the solution space, \eg by choosing a specific algorithm or population size~\cite{zitzler1999multiobjective}.

Optimizing multiple objectives is challenging~\citep{Aleti:2013gp} 
because many refactoring actions improve one or more attributes of the solution space but at the same time they deteriorate other attributes~\citep{SoftwareArchitectureinPractice}, \eg splitting a component improves maintainability but it can also introduce a performance overhead.
As a result, designers must accept trade-offs based on their preferences.

Estimating the quality of Pareto fronts is commonly carried out by exploiting quality indicators~\citep{Li_Yao_2020}.
The purpose of such indicators is to quantify the difference between solution sets.
Each indicator measures a specific aspect of the front compared to some reference point or set. In the following, we describe quality indicators relevant for this work:
\begin{itemize}[noitemsep, nolistsep, leftmargin=*]
    \item Hypervolume (HV)~\citep{Zitzler_Thiele_1998} is arguably the most commonly used quality indicator. HV measures the volume of the search space covered by a front and does not require a reference front. HV especially captures  the quality aspects of convergence (closeness to the reference point), spread (covered region), and cardinality (number of solutions in the set). High values of HV are preferable because they indicate that the front covers a large volume.
    \item Inverted Generational Distance (IGD+)~\citep{Ishibuchi_Masuda_Tanigaki_Nojima_2015} measures the inverted Euclidian distance of a given front from a reference front. Therefore, low values are preferable, and they suggest that the front has good convergence and spread.
    \item Epsilon indicator~\citep{Zitzler_Thiele_Laumanns_Fonseca_da_Fonseca_2003} computes the maximum difference between two fronts. Epsilon is to be minimized and, similarly to IGD+, it provides an estimate of the convergence and spread of a front with respect to the reference front.
\end{itemize}

\subsection{Evolutionary Algorithm}\label{sec:evol-algo}

Genetic Algorithms (GAs) are a class of optimization algorithms that involve several steps to identify the optimal solution(s) to a given problem. 
We briefly describe each step below.

\paragraph*{Initialization} In the first step, an initial population of potential solutions is generated. 
These solutions are typically randomly created within the problem's constraints to cover a broad area of the search space. 

\paragraph*{Evaluation} In this step, each individual within the population is evaluated to determine its fitness using a predefined scoring system. 

\paragraph*{Selection} This process aims to choose individuals who will contribute to the next generation based on their fitness. 
Techniques such as tournament selection, roulette wheel selection, or rank-based selection are used to preferentially select fitter individuals, while maintaining genetic diversity. 

\paragraph*{Reproduction} Crossover and Mutation are genetic operators that generate new offspring by swapping segments of genetic material between two parents to create children, potentially combining beneficial traits from both parents, and introducing random changes to individual genes that help maintaining genetic diversity within the population. 

\paragraph*{Replacement} The offspring generated is integrated into the population through a replacement process, which can be done through generational replacement or incremental strategies like steady-state replacement. 
The algorithm iterates through the cycle of Evaluation, Selection, Crossover and Mutation, and Replacement until a termination condition is met. 

\paragraph*{Termination} The criteria may include reaching a maximum number of generations, achieving a satisfactory fitness level, or observing minimal improvement over several generations. The best solution(s) found during the iterations are presented as the output.

\medskip

\begin{figure}
	\centering
	\includegraphics[width=.5\textwidth]{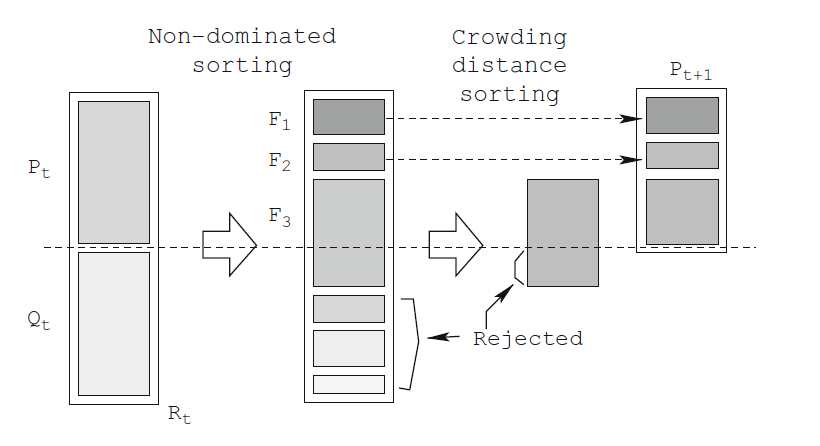}
	\caption{The NSGA-II flow.}
	\label{fig:nsga}
\end{figure}

In this study, we exploited the \nsga algorithm introduced by \citet{DBLP:conf/ppsn/DebAPM00} whose workflow is depicted in \Cref{fig:nsga}.                               
\nsga starts by creating an initial random population ($P_t$). 
The Reproduction and Replacement operators then generate the offspring ($Q_t$). 
The Evaluation step processes the population ($R_t$) and sorts the population members into frontiers ($F_1$ to $F_n$) based on the Non-dominated sorting policy. 
Only members within regions that are less crowded will survive and remain in the population ($P_{t+1}$) for the subsequent optimization step~\citep{DBLP:conf/ppsn/DebAPM00}. 
When the termination criteria are met (e.g., the number of iterations), the final population ($P_{t+1}$) represents the Pareto front generated by the algorithm.

 \subsection{The evolutionary approach}\label{sec:easier}

The optimization process we envision in our study was illustrated in \Cref{fig:schema}.
Such framework starts with an \emph{(initial) architecture model} and a set of \emph{objectives}, then, a \emph{multi-objective optimization engine} is put in place to generate architecture model alternatives. A \emph{visualization} environment is proposed for the designer to visually inspect the solution space.
The designer can select specific regions of the space to focus on, or change the configuration settings of the optimization engine.

The optimization engine is in charge of generating architecture model alternatives through generative operators, such as \emph{selection}, \emph{mutation}, and \emph{crossover}.
In our scenario, such operators apply a fixed number of refactoring actions.
We use the four refactoring actions presented by \citet{CORTELLESSA2023107159} and briefly described in \Cref{tab:ref_actions}.
The correctness of refactoring actions concatenation in sequences is guaranteed by a feasibility engine provided by \easier~\citep{Arcelli:2018vo}.
\begin{table}[]
    \centering
    \begin{tabular}{lp{12cm}}
        \toprule
         Action & Description \\
        \midrule 
         ReDe & Redeploy an existing component on a new node. This action modifies the deployment view by redeploying a Component to a newly created Node. 
         The new Node is connected with all other ones directly connected to the Node on which the target Component was originally deployed. \\
         \midrule
         MO2C & Move an existing operation to an existing component. This action is in charge of selecting and transferring an operation to an arbitrary existing target component. The action consequently modifies each scenario in which the operation is involved. \\
         \midrule
         Clon & Clone a node. This action is aimed at introducing a replica of a Node. Adding a replica means that every deployed artifact and every connection of the original Node has to be, in turn, cloned. \\
         \midrule
         MO2N & Move an existing operation to a new component deployed on a new node. This action is in charge of selecting an operation and moving it to a new component. This action has to synchronize dynamic and deployment views. A lifeline for the newly created component is added in the dynamic view, and messages related to the moved operation are forwarded to it. In the deployment view, a new node, artifact, and related links are created instead.\\
        \bottomrule 
    \end{tabular}
    \caption{The available refactoring actions.}
    \label{tab:ref_actions}
\end{table}
The four objectives are detailed in the following paragraphs.

\paragraph{Performance Quality Indicator (\perfq)}\label{sec:background:perfq}

\perfq quantifies the performance improvement/detriment between two architectural models, and it is defined as follows:
\[perfQ(M)=\frac{1}{c}\sum\limits_{j=1}^{c} p_j\cdot \frac{F_j-I_j}{F_j+I_j}\]
Let $M$ be an architectural model resulting from a sequence of refactoring actions applied to the initial one. 
The performance of $M$ is measured using performance indices ($c$). 
Each performance index is denoted as $F_j$ and represents values such as response time or throughput. 
Additionally, $I_j$ represents the value of the same index on the initial model.

The multiplying factor $p \in \{-1, 1\}$ is used to indicate the objective for the $j$-th index:
(i) When $p=1$, the index is to be maximized, meaning higher values correspond to better performance (\eg throughput).
(ii) When $p=-1$, the index is to be minimized, implying that smaller values indicate better performance (\eg response time).

Finally, the global \perfq is computed as the average across the number $c$ of performance indices considered in the performance analysis.

\paragraph{Refactoring cost (\achanges)}\label{sec:background:distance}

The refactoring cost~\citep{Arcelli:2018vo}, denoted as \achanges in this study, refers to the effort required to transform the initial architectural model into a changed version by applying refactoring actions.
In a previous paper~\cite{CORTELLESSA2023107159}, we introduced the \emph{baseline refactoring factor (\brf)} and the \emph{architectural weight (AW)} metrics to measure the refactoring cost required to apply the actions.
On the one hand, BRF is action-related, and it expresses the effort required to apply a particular action without considering the model element on which the action will be applied.
On the other hand, the AW is element-related, and it expresses the effort required to apply an action to a specific model element.

It is important to note that, in our optimization problem, the relative ratios of \brf values are more crucial than the specific values themselves to prevent the optimizer from being biased towards specific refactoring action types.
The effort needed to perform a refactoring action is quantified as the product of the \emph{baseline refactoring factor} of the action and the \emph{architectural weight} of the model element on which the action is applied.
Finally, the total refactoring cost (\achanges) is obtained by summing the efforts of all refactoring actions contained in a sequence, as shown in the following equation:
\[\achanges(\mathbb{A}) = \sum_{a_i(el_j) \in \mathbb{A}} \brf(a_i) \times AW(el_j)\]

\paragraph{Performance Antipatterns (\pas)}\label{sec:background:pas}

A performance antipattern describes bad design practices that might lead to performance degradation in a system. 
Smith and Williams have introduced the concepts of performance antipatterns in~\citep{DBLP:conf/cmg/SmithW01a,DBLP:books/sp/03/SmithW03}. 
These textual descriptions were later translated into first-order logic (FOL) equations~\citep{DBLP:journals/sosym/CortellessaMT14}.
The evolutionary approach exploits the FOL equations to detect the occurrence of performance antipatterns in the model.
\Cref{tab:supported-pas} lists the performance antipatterns detectable by the optimization engine.

\begin{table*}[htbp]
    \centering
   \begin{tabular}{p{.32\textwidth}p{.63\textwidth}}
    \toprule
        Performance antipattern & Description \\
    \midrule
        Pipe and Filter              & It occurs when the slowest filter in a ``pipe and filter'' causes the system to have unacceptable throughput. \\
	\midrule
        Blob                         & It occurs when a single component either (i) performs the greatest part of the work of a software system or (ii) holds the greatest part of the data of the software system. Either manifestation results in excessive message traffic that may degrade performance. \\
	\midrule
        Concurrent Processing System & It occurs when processing cannot make use of available processors. \\
	\midrule
        Extensive Processing         & It occurs when extensive processing in general impedes overall response time.\\ 
	\midrule
        Empty Semi-Truck             & It occurs when an excessive number of requests is required to perform a task. It may be due to inefficient use of available bandwidth, an inefficient interface, or both. \\
	\midrule
        The Tower of Babel               & It occurs when processes use different data formats and spend too much time converting them to an internal format. \\
    \bottomrule
    \end{tabular}
	\caption{Detectable performance antipatterns in our approach. Left column lists performance antipattern names, while the right column lists performance antipattern descriptions~\citep{Smith:2003wv}.}   
    \label{tab:supported-pas}
\end{table*}

\paragraph{Reliability model (\reliability)}

The reliability parameters are annotated on UML models by means of the MARTE-DAM profile.
The probability of executing a scenario ($p_j$) is specified by annotating UML Use Cases with the \emph{GaScenario} stereotype. 
This stereotype has a tag named \emph{root} that is a reference to the first \emph{GaStep} in a sequence. 
We use the \emph{GaScenario.root} tag to point to the triggering UML Message of a Sequence Diagram and the \emph{GaStep.prob} to set the execution probability.
Failure probabilities of components ($\theta_i$) are defined by applying the \emph{DaComponent} stereotype on each UML Component and by setting, in the \emph{failure} tag, a \emph{DaFailure} element with the failure probability specified in the \emph{occurrenceProb} tag.
Analogously, failure probabilities of links ($\psi_{l}$) are defined in the \emph{occurrenceProb} tag of the \emph{DaConnector} stereotype that we apply to UML CommunicationPath elements. 
Such elements represent the connection links between UML Nodes in a Deployment Diagram.
Sequence Diagrams are traversed to obtain the number of invocations of a component $i$ in a scenario $j$ (denoted by $InvNr_{ij}$ in our reliability model), but also to compute the total size of messages passing over a link $l$ in a scenario $j$ (denoted by $MsgSize(l,j)$). 
The size of a single UML Message is annotated using the \emph{GaStep.msgSize} tag. 
Thus, the mean failure probability $\theta_S$ of a software system $S$ is defined by the following equation:

\[ \theta_S = 1 - \sum\limits_{j=1}^K p_j \left( \prod\limits_{i=1}^N (1 - \theta_i)^{InvNr_{ij}} \cdot \prod\limits_{l=1}^L (1 - \psi_{l})^{MsgSize(l,j)} \right) \]

The Java implementation of the reliability model is available online.\footnote{\url{https://github.com/SEALABQualityGroup/uml-reliability}}

\bigskip 

We remark that we primarily selected the objectives mentioned above because current literature ~\cite{DBLP:conf/icsa/BuschFK19,Aleti:2013gp,CORTELLESSA2023107159} shows that they are important not only in the software architecture domain but also because they create an interesting trade-off space, as they mainly compete against each other.

\section{Related work}\label{sec:related-work}

Steering the search towards regions of the solution space that are of interest (or preferred) to the human architect is still a challenge~\citep{aljawawdehMetaheuristicDesignPattern2015, colanzi2020symposium}. The optimization might spend a considerable time looking for candidate solutions in some space regions that only contain a few relevant solutions (for the architect) and disregard other promising regions, which is not cost-effective.    
    Human interaction can help to steer the search toward the direction desired by the architect, avoiding wasting resources on exploring areas that are not considered interesting~\citep{ramirez2018interactive}.

    There is a trend of adding interaction to architecture optimization by allowing human architects to participate at specific points of the search process~\cite{Vathsavayi2013}\,---\, also known as \textit{human-in-the-loop}. For instance, the user can: \textit{(i)}~judge if specific solutions meet the quality-attribute goals, \textit{(ii)}~provide positive or negative feedback about particular trade-offs (in a multi-objective context), \textit{(iii)}~adjust parameters of the algorithms, \textit{(iv)}~or ask for a more detailed search (\ie exploitation) of solutions following a predefined pattern, among others.
    \citet{Ramirez_Romero_Simons_2019} investigate the state-of-the-art in interactive search-based software engineering and distinguish four categories of interactions. The \textit{preference-based interactivity} comprises interactions the architect uses to express preferences during the search, \eg selecting candidates. Using \textit{interactive re-optimization}, the architect can redefine the search objectives, \eg by removing an objective. Furthermore, in \textit{human-based evaluation}, the architect (partially) inspects and evaluates the candidates, while his/her actions directly impact the generated candidates in a \textit{human-guided search}.
Interactive mechanisms have been employed in search-based optimization tools for other domains, e.g., Marculescu et al.~\cite{marculescu2015initial} investigated the influence of interactions on their search-based testing approach and concluded that interactivity can direct the search into desirable areas of the solution space.
    However, in a structured literature review investigating state of the art until 2019, \citet{Ramirez_Romero_Simons_2019} do not report any approaches specifically for software architecture optimization that use preference-based interactivity.
    
Among more recent works, we identified some that use techniques similar to our proposed framework and/or investigate the effects of human interactions on the results.
The approach by \citet{rebai2020enabling} clusters solutions regarding quality and code locations before developers express their preferences for a few (representative) solutions.
    In their evaluation, they investigated the effect on the proposed solutions' usefulness, execution times, recommended refactorings, and numbers of interactions and compared it to four other (non-)interactive approaches.
In contrast to our work, these authors focus on the QMOOD~\cite{qmood} qualities (flexibility, reusability, understandability, effectiveness, functionality, extendibility) and conduct a user study that does not investigate the quality of Pareto fronts and architectural objectives.

The approach by \citet{alizadeh2020} aims to maximize the QMOOD~\cite{qmood} qualities while minimizing the deviation from an initial solution. It allows users to approve, modify, and reject refactoring actions that lead to solutions and takes this feedback into account. The evaluation compares the approach against other (non-)interactive approaches regarding solution correctness, solution relevance, code-smell fixing capabilities, and achieved quality improvement.

The Nautilus framework for interactive SBSE by \citet{ferreira2020nautilus} allows humans to guide the search to solve various software engineering problems by categorizing solutions as preferred or not preferred.
    While Nautilus can be used to optimize software architectures, it does not provide specific support for this use case. Furthermore, its evaluation mainly focuses on usability rather than on solution quality.
    
Using the approach by \citet{ramirez2018interactive},
    users are presented with solutions obtained from a clustering process and have to reward or penalize aspects of the given solutions. The approach combines qualitative and quantitative evaluation criteria in the objective function, \ie it considers the solution quality and architectural aspects.
    In their evaluation, the authors investigate the impact of the interactions on the solution based on questionnaires and log files. In particular, they investigate the participants' use of the implemented architectural preferences.
    
\citet{chen2022} conducted an empirical study to investigate whether and when performance aspirations (\ie the information in performance requirements that makes it quantifiable) should be considered in configuration tuning. In contrast to our work, they do not consider interactions and focus solely on performance requirements.

    Although our work does not specifically investigate visualization, there are works~\cite{Kudo2012, Schulz2018} on using visualization techniques to support human interaction in the optimization process. These tools allow users to inspect samples of the solution space (\eg via a Pareto diagram) and adjust their search criteria accordingly.
A representative example is the GATSE  project~\cite{GATSE:Procter2019}, a visual prototype for AADL (Architecture Analysis and Description Language) that displays quality-attribute analyses of architectural alternatives, enabling the architect to focus on regions of the quality-attribute space and narrow down the search\,---\ also referred to as “design by shopping.”

While several of the mentioned works include and investigate human interaction in search-based software engineering, our work is the first to investigate the effects of clustering-based candidate selection on the quality of Pareto fronts and architectural models. In particular, our work is the first that does this for the four quality attributes \perfq, \reliability, \achanges, and \textit{performance antipatterns} (\pas).

 \section{Study Design}\label{sec:design}

This study investigates the impact of the designer's interaction on the otherwise fully automated optimization of software architectures.
To this extent, we designed a process in which a human can intervene at specific search points using the \easier optimization engine\footnote{http://sealabtools.di.univaq.it/EASIER/}.
We intend to provide a human-in-the-loop perspective \cite{Vathsavayi2013} that enables designers to contribute to the optimization by expressing preferences about regions of interest within the solution space.
We hypothesize whether the designer's interactions avoid unnecessary search efforts, and, therefore, the extra budget could be spent searching relevant regions, eventually leading to more desirable architectural models.

\subsection{Our Interactive Process}\label{sec:interactive-process}

The interactive process that we have designed has been conceived to be configurable, that is, a designer has the freedom to modify the values of several parameters to make the process more effective in specific contexts. In this section, we identify the parameters that determine a configuration, and provide the default values that we used in the configuration adopted for our experiments.

To support the designer's interactions, the solutions are grouped into \textit{clusters} \cite{Xu2015ACS:Clustering}, each capturing a different trade-off, and labels are automatically assigned to these clusters. 
To assign labels, we apply a 5-point discretization procedure and obtain an ordinal scale for the range of values of each objective. 
For instance, the \reliability values are converted to the ordinal scale \textit{[unreliable, minimally-reliable, average, reliable, very-reliable]}. This scale provides a good separation of the ranges of objective values and permits a simpler interpretation than the original numeric values, particularly for comparison purposes\footnote{The granularity of the discretization for each objective can be configured differently by the designer. Furthermore, other discretization strategies such as quantiles can be alternatively used.}.
A similar procedure is applied to all objectives.
Thus, the label associated with a solution is the combination of the scale values for each objective.
Furthermore, the designer can get a summary of the solution space using prototypes -- called \textit{centroids} for each cluster, where a centroid is a solution (and its associated label) that best characterizes the instances belonging to a cluster.

In our study, we rely on \textit{k-medoids} \cite{Xu2015ACS:Clustering} to determine groups of homogeneous instances in the solution space, which expose potential regions of interest for the designer.
An instance is a vector of objective values for a given architectural model.
The similarity among instances is computed as the Euclidean distance between their corresponding vectors, where the values are standardized before clustering.
The \textit{medoid} (centroid) of a cluster is the instance whose average of dissimilarities to the instances belonging to that cluster is minimal.
The quality of the resulting clusters is assessed with the \textit{silhouette} coefficient, which is a standard metric for assessing clustering quality.

Among the possible types of user interactions, according to \cite{Ramirez_Romero_Simons_2019}, our process supports the \emph{preference-based interactivity}. This type enables humans to guide the search to particular regions of the solution space. 
We have opted for this type of interaction for the following reasons: (i) it drives the process without distorting the nature of the optimization problem, as for example the interaction based on objective elimination does; (ii) its impact on the optimization process can be more or less evident, depending on the context and the decisions taken, hence it enables different levels of human influence that can be compared and evaluated, as we will show in our experimentation. However, with an implementation effort that depends on the complexity of the technique, other types of interactions can be plugged into our process without modifying its structure.

In particular, our process maps the preference to the cluster centroids, in the sense that each centroid represents a trade-off preference for the solution space.
Thus, the designer can select a centroid to be considered as the initial architecture for a subsequent round of optimization.

\begin{figure*}
    \centering
    \includegraphics[width=.9\linewidth]{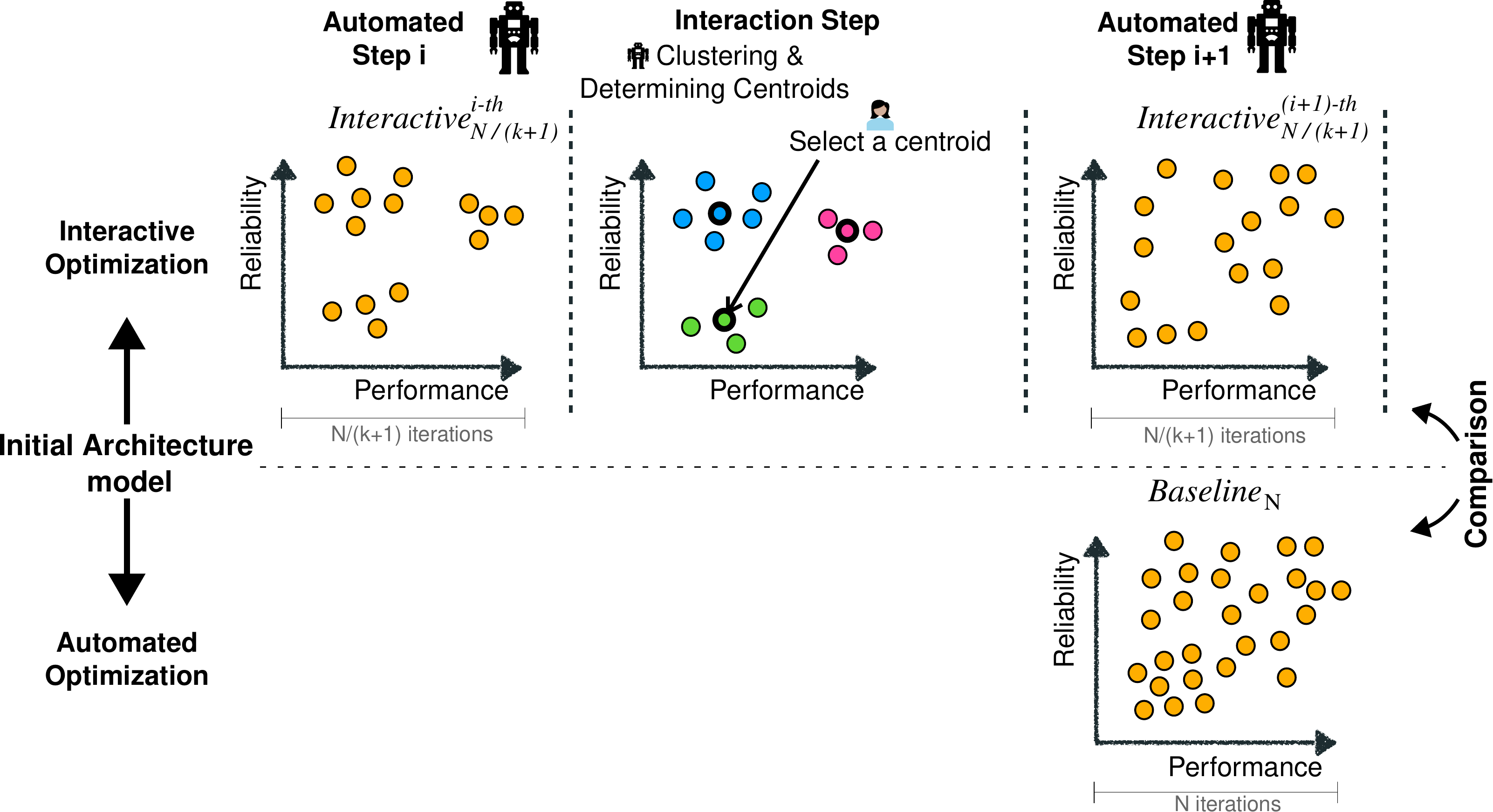}
    \caption{Experimental procedure to compare the interactive and automated processes. For the sake of clarity, we show a two-dimensional optimization space, while the approach supports a four-dimensional space (also including cost and performance antipatterns).}
    
    \label{fig:process}
\end{figure*}

On top of this interactive process, we devise a set of experiments to compare the interactive process to the automated one and assess the quality of the resulting solutions.
This experimental procedure is outlined in \Cref{fig:process}.

We define $Baseline_{N}$ as the automated optimization lasting $N$ evolutionary iterations without interactions, where $N$ can be differently set in different configurations. In the interactive process, several additional configuration parameters are introduced. First, the number of user interactions $k$ has to be defined. This parameter is related to the length $L$ of the sequence of refactoring actions (i.e., the length of the optimization solution chromosome). Indeed, the ratio $L/(k+1)$ determines the number of refactoring actions that are automatically found by the optimization process before each interaction with the user. Analogously, $N/(k+1)$ determines the number of automated process iterations executed before each interaction.

As an example, in our user study we adopt $N=100$ number of iterations, $L=8$ refactoring actions and $k=3$ user interactions. Hence, each automated step executes $(N/(k+1))=25$ iterations to find $(L/(k+1))=2$ refactoring actions (\ie $Interactive^{i-th}_{N/(k+1)}$ in \Cref{fig:process}). Then, the process let the designer pick a centroid, and thereafter the next automated step is executed (i.e., $Interactive^{(i+1)-th}_{N/(k+1)}$ in \Cref{fig:process}) up to the process termination. In practice, in each interaction the designer receives a list of centroids as input, selects the one that is closer to her goals (e.g., high reliability and low cost), and restarts the searching process. \Cref{ssec:user-study-design} provides more details on interactions by describing how users have operated in our user study.

Finally, we also ran a search of $1000$ evolutionary iterations (\refpf) solely to build a reference Pareto front that we use to compute reference quality indicators (\textit{HV}, \igd and \eps, in our case)\footnote{We do not explicit include this number of iterations in the set of configuration parameters, because it is not directly related to the human interactions. Of course, it can be modified on a case basis, mostly depending on the size, spreadness and sparsity of the solution space.}.

\subsection{Research Questions}\label{sec:rqs}

On the basis of the interactive process described in the previous sections, we define the following research questions.

\bigskip
\paragraph*{$RQ1$: To what extent can an interactive approach affect the quality of solutions?\\\\}

To assess human interactions' impact on the generated solutions, we compare the Pareto fronts obtained through the interactive experiments with those resulting from non-interactive runs.
Depending on the problem, several quality aspects can be of interest~\citep{Wu_Arcaini_Yue_Ali_Zhang_2022}, and multiple indicators should be employed to assess quality, because a solution set can be superior to others in terms of one aspect (e.g., coverage) but inferior with respect to other aspects (e.g., diversity)~\citep{Li_Yao_2020}.
Among the many quality indicators available, we selected \ac{HV}, \ac{IGD+}, and \ac{EP} for our study~(see \Cref{sec:bg:sbse}).

HV is usually computed against a reference point, but there is a lack of consensus on how to choose such a point for a given problem.
Here, as a reference point, we use the nadir point, namely the point originated by worst objective values in the Pareto optimal front, because this can be estimated for our problem and it is a common choice in the literature~\cite{Li_Yao_2020}.

While being helpful in establishing a quantifiable relation between fronts, quality indicators can be challenging to connect to a comprehensive view of the solutions. In order to make the best out of the estimates provided by quality indicators, we associate them with the software attributes that a designer is trying to improve. Therefore, this research question investigates how quality indicators relate to the quality of solutions in terms of performance and reliability, which are critical quality attributes in our systems under test.

 \bigskip
\paragraph*{$RQ2$: How different are, in terms of architectural properties, the solutions generated through an interactive process with respect to the ones generated through a fully automated one?\\\\}

We are interested in studying specific characteristics of solutions that concern their architectural aspects and go beyond their quality attribute metrics (i.e., performance and reliability).  

At first glance, the architectural models can be analyzed in terms of cost of refactoring (\achanges) the initial reference architecture towards the selected target design, and performance antipatterns present in the architecture (\pas), as these metrics quantify the structural characteristics of the architectural models. 

To better understand the possible variations of \achanges and \pas, we partition the ranges of values for these objectives into five bins (or categories) and derive a 5-point ordinal scale for each objective. For instance, for \achanges, the scale goes from \textit{very-few} (lowest, best value) to \textit{many} (highest, worst value). A similar scale is applied to \pas.

To analyze the models in more detail, we look at the refactoring actions contained in the solutions. In particular, we consider the types of refactoring actions (see Table $1$) and their arrangement into sequences, as they provide insights into the constructive patterns resulting from the optimization process. We use a given sequence $S=<ra_1,ra_2,...,ra_n>$ as a proxy for the set of architectural models resulting from applying refactoring actions $ra_1,ra_2,...,ra_n$ (with different parameters) on the initial architecture. 
To assess whether the sequences of refactoring actions from different experiments differ, we compute their intersection. 

For a given experiment, we analyze the frequency of the action types used across the sequences. In particular, in a non-interactive process, the possible actions to appear in a sequence are unconstrained. However, when the designer selects an intermediate solution, this sets the initial actions for the sequences to be generated in subsequent optimization runs. We then assume that not all the refactoring actions are used evenly in the sequences. The frequent usage of certain action types in the sequences could correlate with particular architectural models.

 \bigskip
\paragraph*{$RQ3$: How does an interactive process impact the coverage of the solution space?\\\\}

In this research question we look at the whole four-dimensional solution space. The distribution of the objective values, or landscape, might spread differently or exhibit areas with a higher (or lower) concentration of solutions, depending on whether the designer's interactions are considered.

To visually characterize the landscapes, we employ kernel density estimation (KDE) functions based on heatmaps. A KDE is a mathematical process for computing the probability density function of a random variable in order to create a smooth curve (or surface) for a finite data set (\eg a scatter plot). The contribution of each point is smoothed out from a single point into a neighborhood. 
Since our landscapes are four-dimensional, we first apply a Principal Component Analysis (PCA) transformation of the objectives to a 2D space. PCA is a dimensionality reduction technique that projects the data (originally in a high-dimensional space) onto a set of orthogonal axes (in a lower-dimensional space) in such a way that most variance of the data is maintained. Although the 2D PCA projections are approximations, the KDE heatmaps allow us to qualitatively compare the landscapes resulting from each experiment.

In addition, we rely on an entropy-based metric \cite{Farhang:2004} to quantify the objective trade-offs over the landscape. Entropy is computed on a four-dimensional histogram of the space. The histogram bins represent trade-off categories of solutions, and they are derived from the 5-point discretization scale mentioned in \textit{RQ2}, while considering all possible combinations of the objectives. For example, the bin that contains the best optimization values (for \perfq, \reliability, \achanges, and \pas, respectively) is \textit{<very-fast, very-reliable, very-few, very-few>}. The histogram has, in principle, $5^4$ bins of possible trade-offs, but in an experiment not all of them might contain solutions. 
The entropy-based metric gets closer to $1$ if the space density is primarily flat and homogeneous (representing a non-informative landscape), and it tends to $0$ when the space includes peaks and becomes heterogeneous (representing an informative landscape) that is the most interesting scenario for the designer. 

For \textit{RQ3}, we arrange all the sequences of refactoring actions of an experiment as a tree, in which the leaves correspond to architectural models and the inner nodes capture refactoring actions shared by the different sequences. 
The tree representation allows the designer to identify the order of actions applied to each specific architectural model and determine if the spaces explored from two experiments have intersections (i.e., they share common subtrees). The designer can then use these insights to determine the specific regions of interest based on the quality of solutions.
In particular, we build a reference tree $T_R$ for the \refpf and then map the trees resulting from the three experiments (the \base and the two interactive ones) to $T_R$. The intersection of the trees is an estimation of how much of the search space (as represented by \refpf) is covered by the interactive and non-interactive experiments. The coverage of any tree $T_i$ over $T_R$ is computed as the proportion of nodes in $T_R$ that also appear in $T_i$.

Additionally, we compute the coverage metric~\cite{zitzler1999multiobjective} to quantify the percentage of the \base Pareto front that is covered by the interactive processes, and vice versa. The coverage metric of two sets of solutions $X^{\prime}$ and $X^{\prime\prime}$ is defined as:
\begin{equation}
    \text{C}(X^{\prime}, X^{\prime\prime}) = \frac{|a^{\prime\prime} \in X^{\prime\prime}; \exists a^{\prime} \in X^{\prime} : a^{\prime} \preceq a^{\prime\prime}|}{|X^{\prime\prime}|}.
\end{equation}
The coverage metric ranges from $0$ to $1$, where $1$ indicates that all solutions in $X^{\prime\prime}$ are covered by (\ie dominated by or equal to) $X^{\prime}$, and $0$ indicates that none of the solutions in $X^{\prime\prime}$ are covered by $X^{\prime}$.

 \subsection{Benchmark Systems}\label{sec:case-study}

We faced a challenge in finding systems complex enough for our study.
        To tackle this issue, we selected two significant studies - \textit{Train Ticket Booking Service} (\ttbs)~\cite{DBLP:conf/staf/Pompeo0CE19} and \textit{CoCoME}~\cite{Herold2008} - and adapted them to our needs.
        The \ttbs model was derived from a microservice-based application~\cite{DBLP:journals/tse/ZhouPXSJLD21} that served as a reference in many other studies.
        On the other hand, \ccm is a reference system for non-functional model-based analyses.
        We have included the architectural specifications of each system, such as static, dynamic, and deployment diagrams (in UML format), and analytical models (LQN) in the replication package\footnote{\url{https://github.com/SEALABQualityGroup/uml2lqn-casestudies/}}, to make the results easier to be replicated.
It is worth noting that the two benchmark systems have different characteristics, allowing our approach to be experimented in different contexts. 

\paragraph{Train Ticket Booking Service}

\ttbs is a web-based booking application whose architecture is based on the microservices paradigm~\citep{DBLP:journals/tse/ZhouPXSJLD21}. 
The system is made up of $40$ microservices, each one deployed on a Docker container.  Based on the system specification and requirement documents, we specified \expComp{11} in describing the static view, \expNode{11} for the deployment view, and \expUC{3} (\ie \emph{login}, \emph{update user details}, and \emph{rebook}) for the dynamic view of the system.

\paragraph{CoCoME}

\ccm describes a trading system containing several stores. 
A store might have one or more cash desks for processing goodies. 
A cash desk is equipped with all the tools needed to serve a customer. Based on the system specification, we specified a static view with \expComp{13}, a deployment view with \expNode{8}, and a dynamic view with \expUC{3}.
 \subsection{Experimental Setup}\label{sec:exp-setup}

We report the \textit{Easier} algorithm configurations in \Cref{tab:nsga_config}. 

Regarding the number of iterations, the optimization can run for more than 100 iterations and the designer can interact with the optimization more than once, as it occurs in our user study. We determined these values based on whether the evolutionary algorithm converges to Pareto-optimal solutions for the system under test and the number of clusters appearing after $50$ iterations. A larger number of interactions is desirable for a more complex system and a larger number of iterations.

Since an exhaustive tuning of all the parameters and their combinations is not feasible\footnote{Hyperparameter tuning is a common practice in Genetic Algorithms. However, it can be time-consuming, and some parameters in our domain cannot be adjusted automatically due to their specific semantics. Therefore, we manually configured certain parameters and found no issues with the configurations used in this study.
}, we used  \emph{$P_{crossover} = 0.8$} and \emph{$P_{mutation} = 0.2$}, as suggested by \citet{Arcuri_Fraser_2011}. For the operators, we used \emph{Single Point} crossover, \emph{Binary Tournament} selector, and \emph{Simple Mutation} operators, as provided by the JMetal\footnote{\url{https://github.com/jMetal/jMetal}} library.
Furthermore, we set the \emph{Population size} to $16$ elements, and the \emph{length} of the chromosome to $4$ for the \refpf and \base runs, as reported by \citet{Di_Pompeo_Tucci_2022}, and \emph{length} of $2$ for the interactive runs\footnote{
Before starting the optimization, users can select the number of interactions they want, and other parameters such as chromosome length and number of iterations. The framework then divides the iterations and the chromosome into equal segments of duration and chromosome length. For instance, if the user sets one interaction during 100 iterations with a chromosome length of 4, the framework will create two segments of 50 iterations, each with a chromosome length of 2.}.
Each configuration was executed in both benchmark systems 31 times (as suggested by \citet{Arcuri_Briand_2011}).
Our experiments lasted the equivalent of \emph{126} days on three Dell PowerEdge C6525 servers, each equipped with two AMD EPYC 7282 2.80GHz CPUs and 512 GiB of RAM.

Naturally, the time spent on each experiment depends very much on the number of iterations and the population size. As a rough estimate, assuming a single subject system and the settings in \Cref{tab:nsga_config}, the time spent on a single interaction step is around $4$ hours, including the time to compute the Pareto front and the time to cluster the solutions.

\Cref{tab:cs_config} lists the cluster centroids selected by the interactions. Each interaction step involved four possible centroids, which were determined using the \textit{k-medoids} algorithm and looking for the best \textit{silhouette} coefficient. For the first iteration of both \ttbs and \ccm, two authors picked two (out of four) representative centroids each (on the basis of their individual preferences) in order to keep the analysis manageable (that is, to limit the number of branches to explore during the experiment, see Figure~\ref{fig:userstudy-charts}). The centroid attributes include its name (first column), the objectives, and their values (second and third column).  
The centroid label (second column) describes the trade-off of the corresponding cluster (third column).
We released the datasets in our replication package\footnote{\url{https://github.com/SpencerLabAQ/replication-package__interactive-search-based-software-performance}}.

\begin{table}
        \centering
        \begin{tabular}{lp{.65\linewidth}}
                \toprule
                Configuration & Values  \\
                \midrule
		 Number of iterations (N) &  1000 (reference), 100 (baseline), 50 (interaction), 25 (user study) \\
                 Population Size & 16 \\
		 Length (L) & 4 (reference / baseline), 2 (interaction), 8 (user study) \\
		 Number of user interactions (k) & 1, 3 (user study) \\ 
                 Number of independent runs & 31 \\
                 Selection operator & Binary Tournament \\ $P_{crossover}$ & 0.80 \\
                 Crossover Operator & Single Point \\
                 $P_{mutation}$ & 0.20 \\
                 Mutation Operator & Simple Mutation \\
                \bottomrule
        \end{tabular}
	\caption{\label{tab:nsga_config} \nsga configurations. \DDP{I updated the table accordingly to the text in \Cref{sec:interactive-process,sec:exp-setup}}}\end{table}

\begin{table}
        \centering
        \begin{tabular}{llr}
                \toprule
                Centroid & Objective & Value  \\
                \midrule
\multicolumn{3}{c}{\ttbs} \\
                \midrule
		\multirow{5}{*}{c223}  & \perfq          & 0.205996                                  \\
		{}                     & \reliability    & 0.788695                                  \\
		{}                     & \achanges       & 2.9                                       \\
		{}                     & \pas            & 2                                       \\
		{}                     & label           & fast / very-reliable / very-few / average \\
		\midrule
		\multirow{5}{*}{c258}  & \perfq          & 0.208608                                        \\
		{}                     & \reliability    & 0.302645                                        \\
		{}                     & \achanges       & 3.28                                            \\
		{}                     & \pas            & 0                                             \\
		{}                     & label           & fast / minimally-reliable / very-few / very-few \\
		\midrule[1.3pt]
		\multicolumn{3}{c}{\ccm} \\
                \midrule
		\multirow{5}{*}{c317}  & \perfq          & 0.216461                    \\
		{}                     & \reliability    & 0.759612                    \\
		{}                     & \achanges       & 3.28                        \\
		{}                     & \pas            & 10                        \\
		{}                     & label & fast / very-reliable / very-few / few \\
		\midrule
		\multirow{5}{*}{c358}  & \perfq          & -0.014243                             \\
		{}                     & \reliability    & 0.853851                              \\
		{}                     & \achanges       & 3.28                                  \\
		{}                     & \pas            & 8                                   \\
		{}                     & label           & slow / very-reliable / very-few / few \\
                \bottomrule
        \end{tabular}
        \caption{\label{tab:cs_config} Centroids properties. We report centroid properties in terms of objectives and clustering labelings. For the clustering labels, we use a 5-point Likert scale: \emph{performance} / \emph{reliability} / \emph{cost} / \emph{performance antipatterns}.}
\end{table}

\subsection{User Study Design}
\label{ssec:user-study-design}

\ADP{This paragraph is mentioned almost exactly in the Introduction. I rephrased it a bit here}In the previous research questions, we examined the measurable effects of the interactive process. However, its success also depends on designer acceptance. We now aim to explore the human perspective, assessing whether the features of the process are seen as useful and if designers recognize benefits over a fully automated approach.

In order to assess our interactive approach with human subjects, we designed a user study in which we asked a group of software architecture experts to use a tool we designed, implemented with a \textit{Jupyter} notebook, to explore alternative solutions for the \ttbs system. The study consisted of a series of design sessions with $14$ participants, mainly from academia. The participants had different positions (2 assistant professors, 12 Ph.D. students) and varying degrees of knowledge.
An assistant professor had a strong background in software engineering and another one also held a CTO position in an IT company, thus their combination brought both academic and some industry perspectives. The Ph.D. students are enrolled in software engineering and computer science, with varying familiarity in software architecture and multi-objective optimization. One of them, for instance, had software architecture topics as part of his/her doctoral research. However, most participants had limited or no significant industry experience, which suggests their decision-making and design behaviors might align more closely to those of inexperienced software architects.
All sessions had a duration of approximately $45$ minutes and were conducted in person or online. In all sessions, one of the co-authors was present to observe the users and answer their questions.

\begin{figure*}
    \centering
\includegraphics[width=0.98\linewidth]{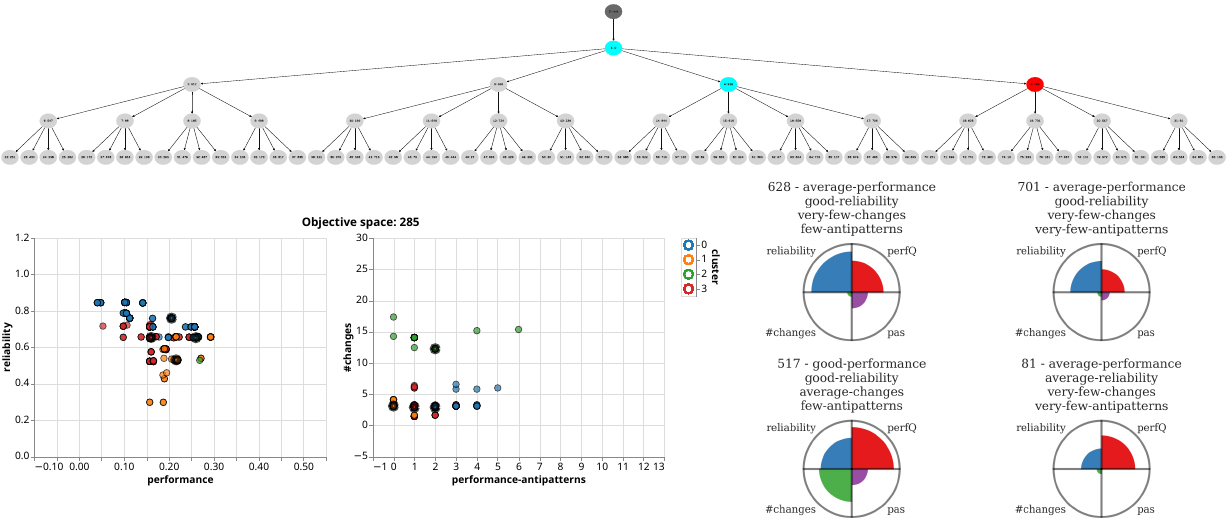}
    \caption{\label{fig:userstudy-charts} Visualizations used during user study sessions.}
\end{figure*}

A session involved four phases, namely: Introduction, Learning, Testing, and Postmortem. The session started with an introduction to the architectural optimization problem under consideration (i.e., \ttbs). The problem consisted of $4^3=64$ possible points (or paths) for user interaction\footnote{Here $4$ is the number of centroids per interaction point, and $3$ is the number of levels in the navigation tree.}, which led to a total of $41,821$ architecture solutions potentially to be explored. Participants were expected to perform successive rounds of search, starting from an initial solution (and its solution space) and selecting centroids according to their own criteria that would lead them to different spaces. For the sake of limiting the duration of the experiment, all potential solutions were computed beforehand and progressively shown to participants as they progressed in their exploration paths. After the introduction, we explained to the participants how to work with the notebook using simplified Python functions and visualization charts. The charts used by the participants during the sessions are depicted in Figure \ref{fig:userstudy-charts}. 
The navigation tree (top) shows the $64$ interaction points that the user could explore. The pairplot of trade-offs (bottom left) shows the solutions and clusters in the space resulting from the current interaction point \textit{NL196} (marked in red in the navigation tree). The radar charts (bottom right) correspond to the $4$ centroids for the clusters in the pairplot, which are themselves interaction points leading to $4$ possible exploration paths in the tree.
In fact, as part of this learning phase, we introduced the notion of clustering to group solutions and identify centroids that lead to different nodes in the navigation tree.

At the end of the learning phase, we launched the exercise itself. In this testing phase, the participant departed from a predetermined (initial) architecture and freely chose the interaction points available in the tree. At each interaction point, the tool automatically displays the charts, and the participant can pick an intermediate solution point (i.e., a centroid) and ask the tool to search for solutions reachable from that point. Participants were able to explore three interaction points (or levels in the tree). As a participant drilled down in the tree to inspect solutions, our tool also allowed them to go back to previous tree nodes (e.g., jumping from an architecture in level $2$ to an architecture in level $1$ in order to investigate alternative interaction points for that level).
We asked participants to take notes about the solutions (tree nodes) visited during the session, particularly those that best reflected the participants' goals or preferences for the exploration. Since all the solutions originate from different trade-offs, we did not intend the participants to optimize for particular objectives. 

In the end, we asked participants to submit a questionnaire to measure their satisfaction levels with both the solutions being explored and the interaction mechanism. To do this, we used a list of Likert-scale questions, which are captured in Figure \ref{fig:results-userstudy}. The questions covered several aspects such as: quality of solutions, comparison to a fully automated approach, role of the interaction mechanism, perceived limitations, and practicality of the approach.  In addition to satisfaction, we also focused on the expectations of the participants as well as areas of improvement for our process, which were gathered as open (textual) feedback. Both the \textit{Jupyter} notebook and the questionnaire are included in our replication package.
 \\

 \section{Results}\label{sec:results}

\subsection{Quality of solutions (RQ1)}\label{sec:results-rq1}

\newcommand{\vda}{$\hat{A}_{12}$\xspace}
\newcommand{\mwu}{Mann--Whitney U\xspace}

\definecolor{LightGray}{gray}{0.95}

Tables \ref{tab:qi-ttbs} and \ref{tab:qi-ccm} report the values of the quality indicators computed for the \ttbs and \ccm systems, respectively.
Here, \refpf represents the best we can achieve in our settings. In the case of \igd and \eps, the value reported for \refpf is zero because these quality indicators are computed against \refpf itself, as explained in \Cref{sec:rqs}.
\base represents the experiments without interaction, against which we intend to compare the interactive ones. 

The first observation is that the cardinality of the Pareto fronts (NPS)\,---\,i.e., the number of solutions in the Pareto front\,---\, is noticeably reduced by the interaction in all cases.
This suggests that the interaction might have the effect of narrowing down the search to specific areas of the solution space, therefore leading to a fewer number of non-dominated solutions.
The rest of the quality indicators seem to agree more or less on how far the interactive experiments are from \base, and that \base itself achieved a Pareto front that is very close to the one obtained by \refpf (see difference in \hv).

For \hv, in the \ttbs case, the interactions covered roughly half of the volume of the \base, whereas, for \ccm, the values obtained covered around 70\% of the \base\xspace (in Table~\ref{tab:qi-ccm}, the values of \hv for \ccsecondintcentrA and \ccsecondintcentrB are $73.8\%$ and $69\%$ of the value of \hv for the \base, respectively).
A similar trend can be observed for \igd and \eps.
This confirms that concerning the solution space, the interaction in \ccm led to solutions being ``closer’’ than \ttbs to those in the \base.
Such a result could be considered unexpected since the solution space of \ccm is larger than that of \ttbs, as reported by \citet{CORTELLESSA2023107159}.
Therefore, it should be easier for optimization to approach the \base in \ttbs compared with \ccm.
On the contrary, it seems that the interaction can drive the search into a narrower space, at least in the \ccm case, but one that contains better solutions and is faster to reach (with 100 iterations).

\begin{table}[htbp]

\centering
\subfloat[\ttbs]{

\begin{tabular}{lrrrr}
  \toprule
  Experiment        & NPS & \hv ($\uparrow$)  & \igd ($\downarrow$) & $EP$\xspace($\downarrow$) \\
  \midrule
  \rowcolor{LightGray}
  \refpf             & 24  &  0.726459 ($\pm 0.041$)  &  0.013483 ($\pm 0.021$)  &  0.204289 ($\pm 0.032$)  \\
  \base              & 14  &  0.684648 ($\pm 0.032$)  &  0.025021 ($\pm 0.025$)  &  0.166667 ($\pm 0.048$)  \\  
\rowcolor{LightGray}
  \ttsecondintcentrA &  2  &  0.392059 ($\pm 0.050$)  &  0.133606 ($\pm 0.050$)  &  0.464450 ($\pm 0.032$)  \\
  \rowcolor{LightGray}
  \ttsecondintcentrB &  2  &  0.344125 ($\pm 0.061$)  &  0.172568 ($\pm 0.029$)  &  0.447109 ($\pm 0.051$)  \\  
  \bottomrule
\end{tabular}

\label{tab:qi-ttbs}

}

\vspace{1em}
\centering
\subfloat[\ccm]{

\begin{tabular}{lrrrr}
  \toprule
  Experiment         & NPS & \hv ($\uparrow$) & \igd ($\downarrow$) & $EP$\xspace($\downarrow$) \\
  \midrule
  \rowcolor{LightGray}
  \refpf             & 89  &  0.904444 ($\pm 0.094$)  &  0.002315 ($\pm 0.048$)  &  0.026316 ($\pm 0.111$)  \\
  \base              & 89  &  0.911122 ($\pm 0.048$)  &  0.003231 ($\pm 0.026$)  &  0.024332 ($\pm 0.070$)  \\  
\rowcolor{LightGray}
  \ccsecondintcentrA & 13  &  0.672427 ($\pm 0.031$)  &  0.061924 ($\pm 0.011$)  &  0.235570 ($\pm 0.042$)  \\
  \rowcolor{LightGray}
  \ccsecondintcentrB & 30  &  0.628814 ($\pm 0.050$)  &  0.075031 ($\pm 0.029$)  &  0.285596 ($\pm 0.058$)  \\
  \bottomrule
\end{tabular}

\label{tab:qi-ccm}
}

\caption{Quality indicators across experiments. NPS is the number of solutions in the Pareto front. Arrows, next to quality indicators, indicate whether high or low values are to be preferred. Standard deviations are shown in parentheses.}
\end{table}
 
\begin{figure}[htbp]
    \centering
    \begin{subfigure}{.9\linewidth}
        \centering
        \includegraphics[width=\linewidth]{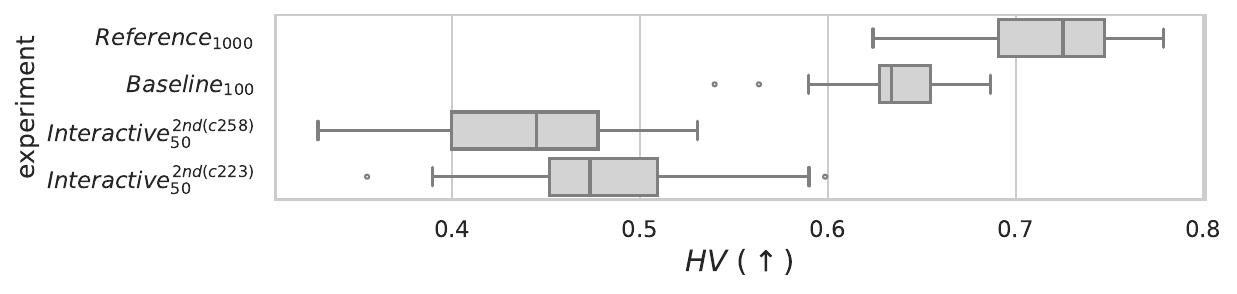}
        \label{fig:rq1-qi-boxplots-ttbs-hv}
    \end{subfigure}
    \par
    \vspace{-1em}
    \begin{subfigure}{.9\linewidth}
        \centering
        \includegraphics[width=\linewidth]{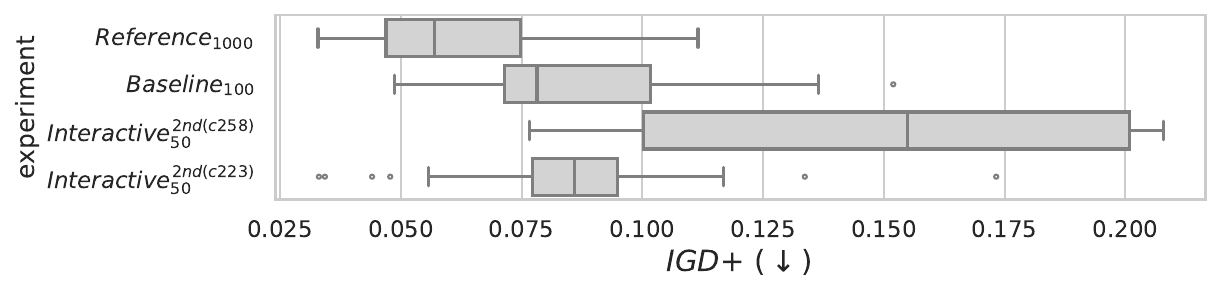}
        \label{fig:rq1-qi-boxplots-ttbs-igd}
    \end{subfigure}
    \par
    \vspace{-1em}
    \begin{subfigure}{.9\linewidth}
        \centering
        \includegraphics[width=\linewidth]{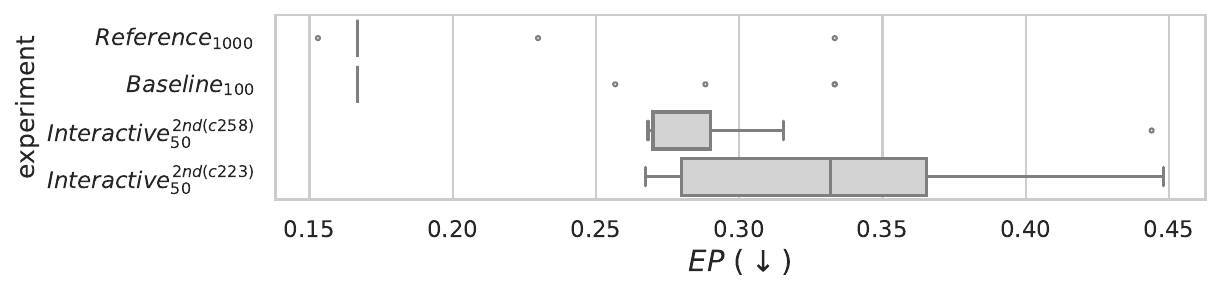}
        \label{fig:rq1-qi-boxplots-ttbs-eps}
    \end{subfigure}
    \caption{Boxplots of the quality indicators for the \ttbs system.}
    \label{fig:rq1-qi-boxplots-ttbs}
\end{figure}

\begin{figure}[htbp]
    \centering
    \begin{subfigure}{.9\linewidth}
        \centering
        \includegraphics[width=\linewidth]{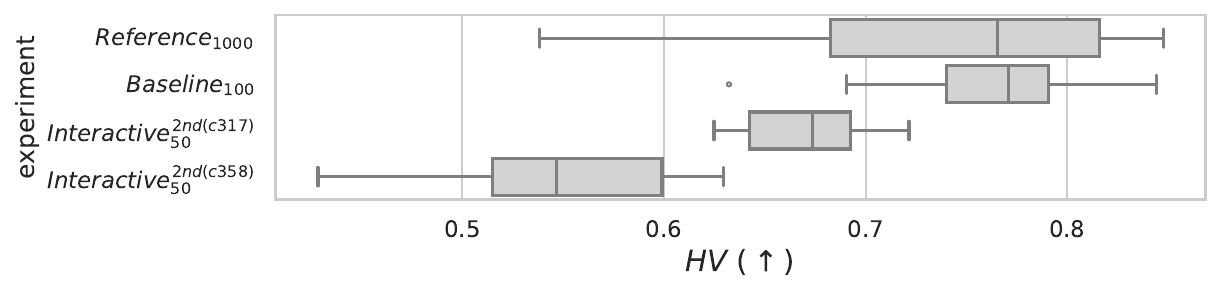}
        \label{fig:rq1-qi-boxplots-ccm-hv}
    \end{subfigure}
    \par
    \vspace{-1em}
    \begin{subfigure}{.9\linewidth}
        \centering
        \includegraphics[width=\linewidth]{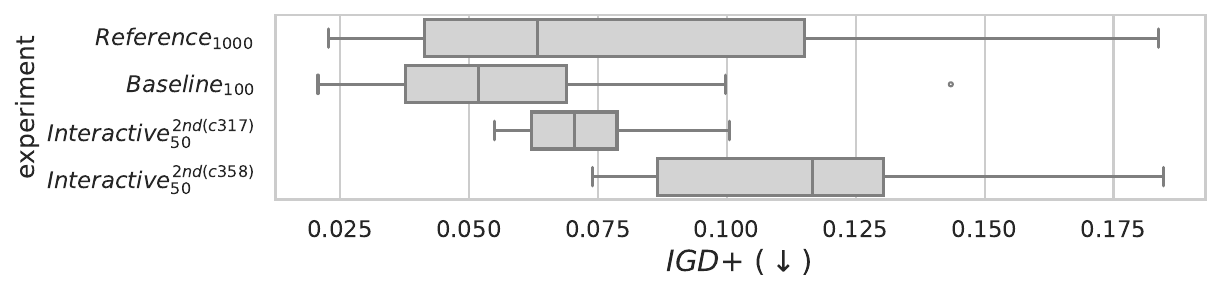}
        \label{fig:rq1-qi-boxplots-ccm-igd}
    \end{subfigure}
    \par
    \vspace{-1em}
    \begin{subfigure}{.9\linewidth}
        \centering
        \includegraphics[width=\linewidth]{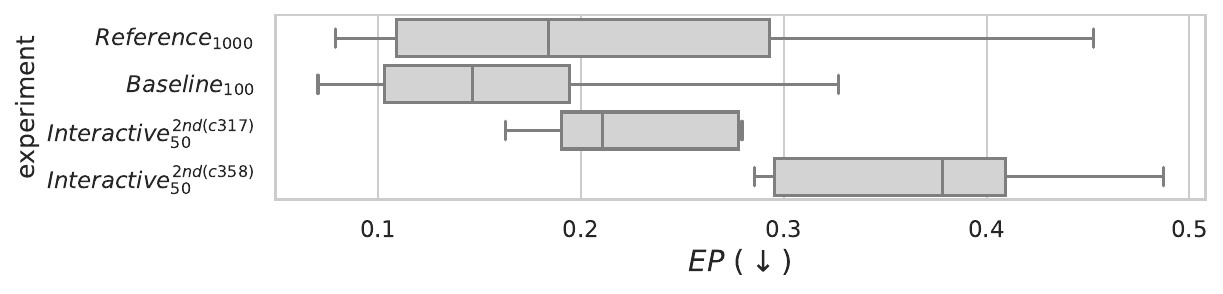}
        \label{fig:rq1-qi-boxplots-ccm-eps}
    \end{subfigure}
    \caption{Boxplots of the quality indicators for the \ccm system.}
    \label{fig:rq1-qi-boxplots-ccm}
\end{figure}

Next, we analyze the variability of quality indicators between runs.
Boxplots in \Cref{fig:rq1-qi-boxplots-ttbs,fig:rq1-qi-boxplots-ccm} show the distribution of the quality indicators for the \ttbs and \ccm systems, respectively.
A different degree of variability can be observed among the quality indicators, regardless of the system, the quality indicator, or the experimental setting.
In some cases, the interactive experiments led to higher variability (\eg \ttsecondintcentrA for \igd in \ttbs), while in others, the variability was reduced (\eg \ccsecondintcentrA for \igd in \ccm).
In general, we cannot conclude whether the interaction reduces or increases the variability of the quality indicators among the runs, as its effect is not consistent across the quality indicators.

An additional analysis was performed to compare the quality indicators of the interactive experiments with the \base, as well as how \perfq and \reliability are distributed in the Pareto fronts.

\begin{figure*}[htbp]
    \centering
    \begin{subfigure}{.45\linewidth}
        \centering
        \includegraphics[width=\linewidth]{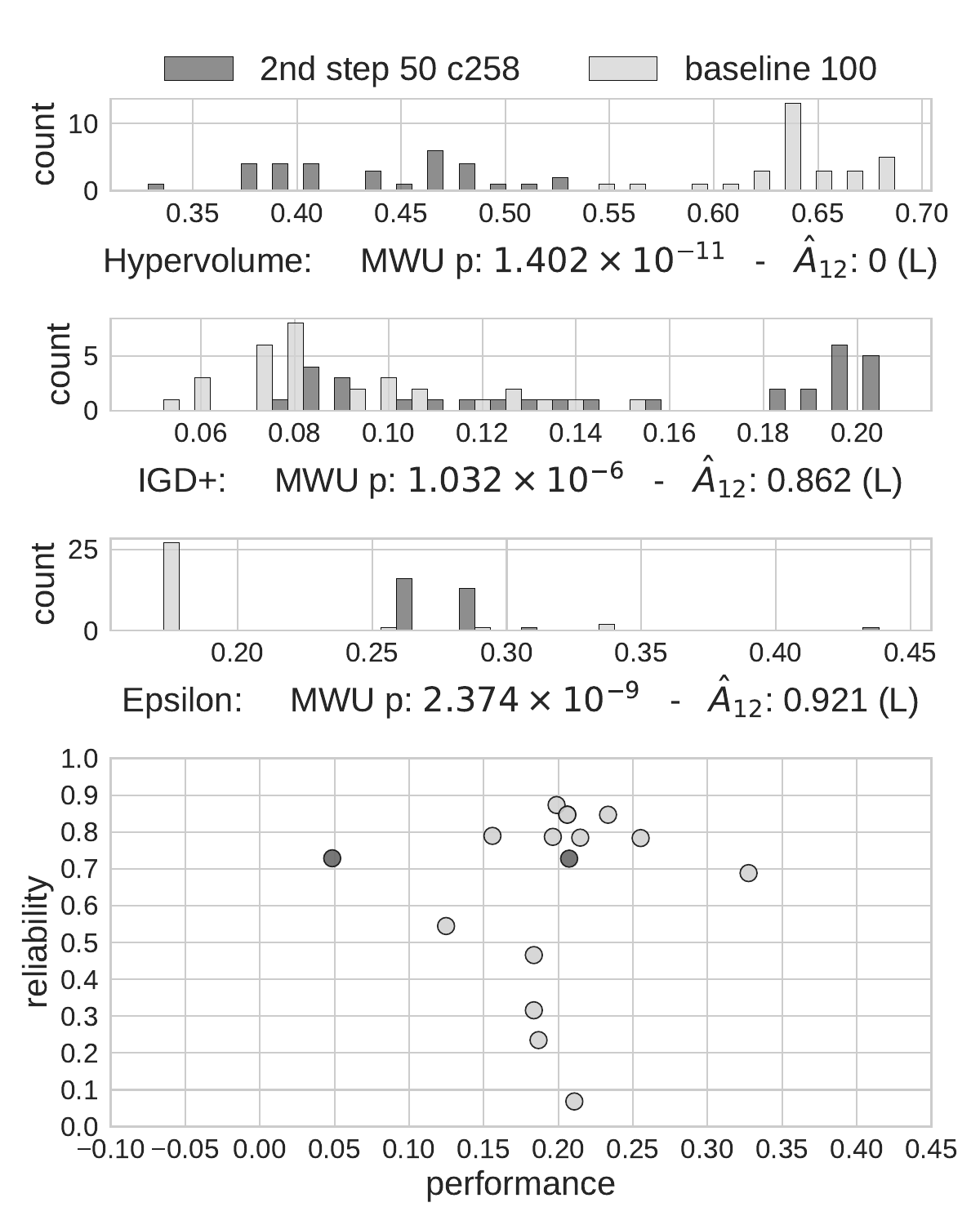}
        \caption{\ttsecondintcentrA against \base}
        \label{fig:rq1-ttbs-inter-centroid-first}
    \end{subfigure}
    \hfill
    \begin{subfigure}{.45\linewidth}
        \centering
        \includegraphics[width=\linewidth]{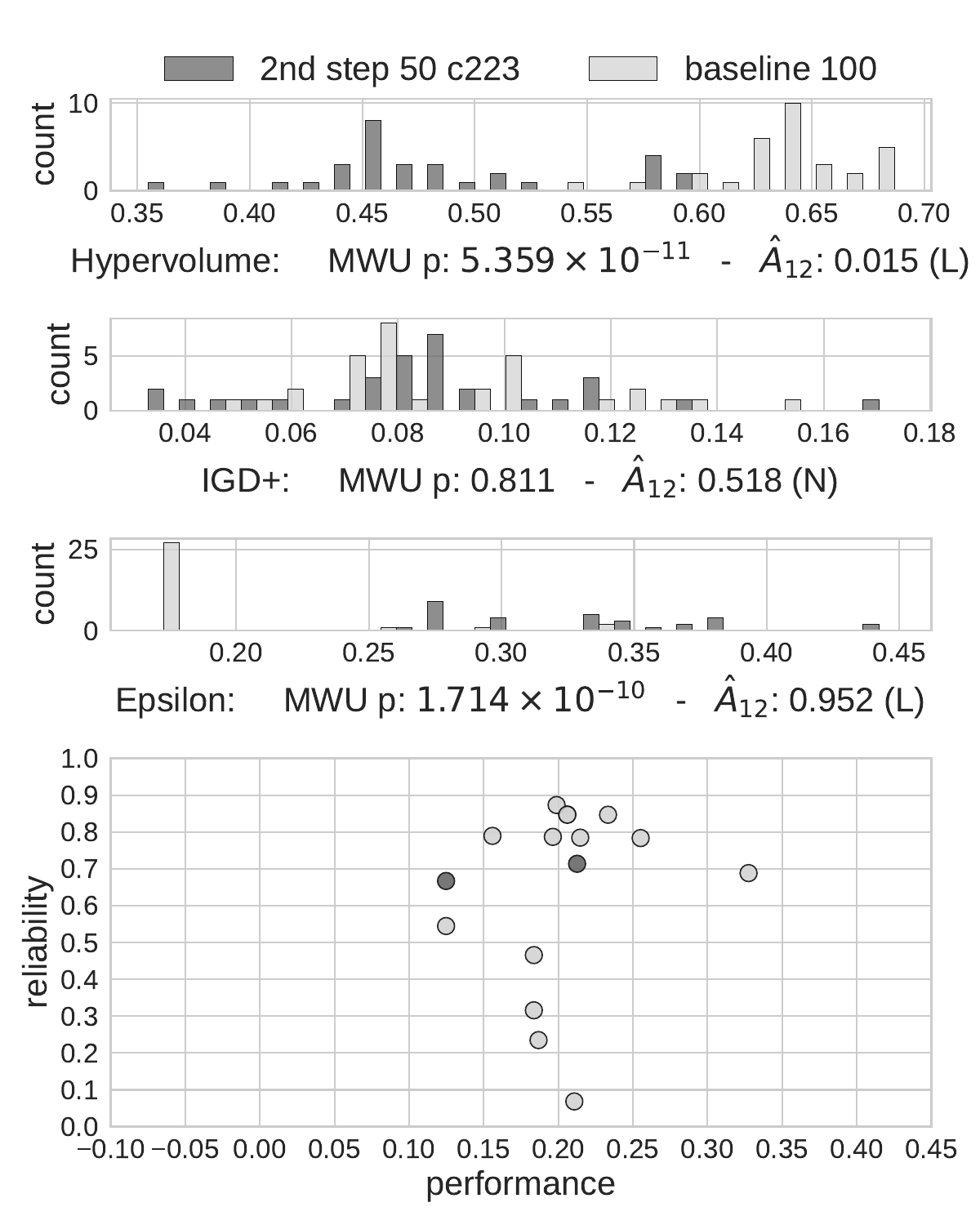}
        \caption{\ttsecondintcentrB against \base}
        \label{fig:rq1-ttbs-inter-centroid-second}
    \end{subfigure}
    \caption{\ttbs: comparison of the interactive steps based on the selection of a cluster centroid against \base, in terms of \perfq, \reliability, and quality indicators. The upper half shows the distribution of the quality indicators among the (31) runs, along with the \textit{p-value} of the \mwu test and the \vda effect size; magnitude interpretation: negligible (N), small (S), medium (M), large (L). The lower half reports the Pareto fronts obtained from both the interaction and \base, projected in a bidimensional space to compare the values of \perfq and \reliability.}
    \label{fig:rq1-ttbs-interactive-centroid-vs-baseline}
\end{figure*}

\begin{figure*}[htbp]
    \centering
    \begin{subfigure}{.49\linewidth}
        \centering
        \includegraphics[width=\linewidth]{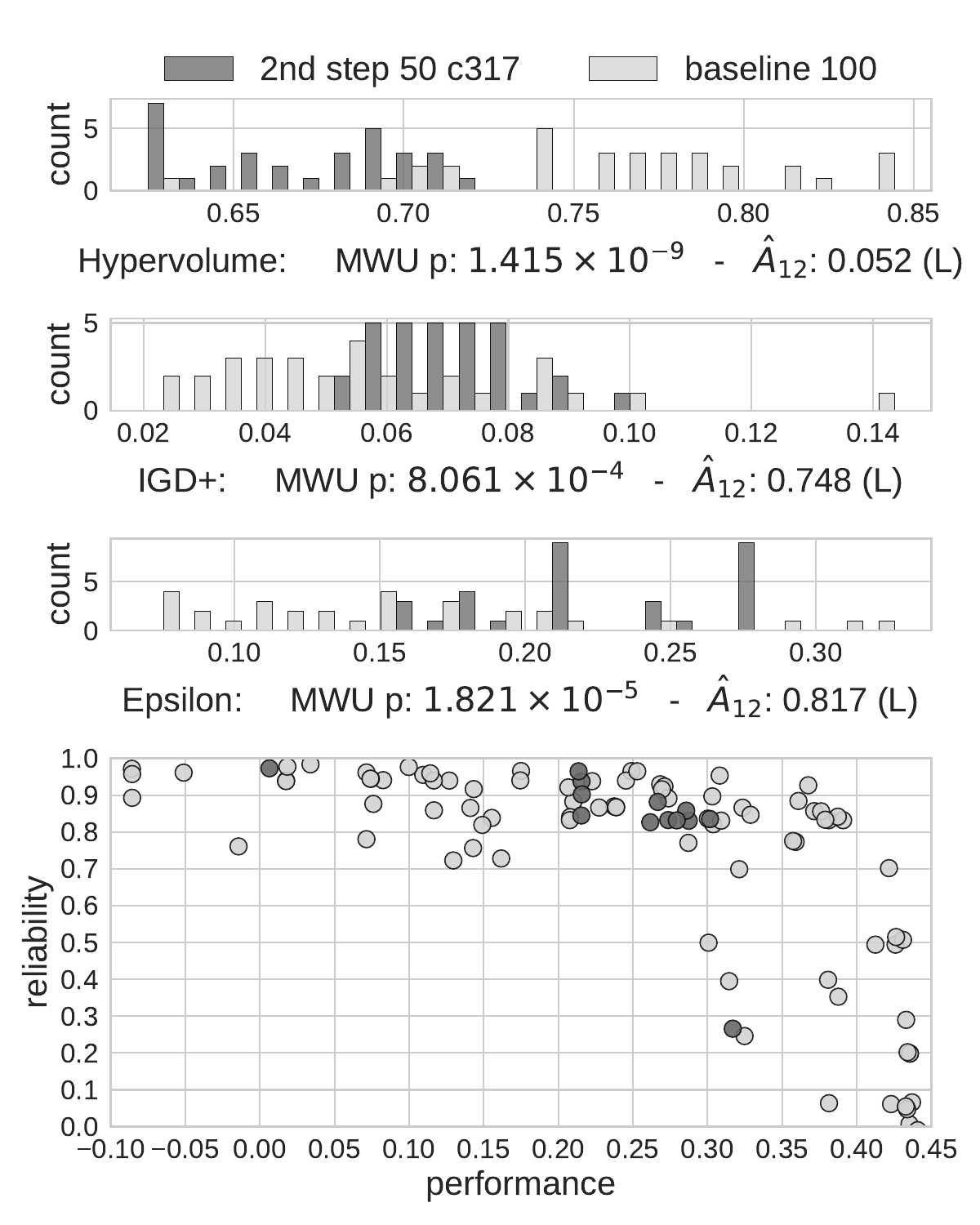}
        \caption{\ccsecondintcentrA against \base}
        \label{fig:rq1-ccm-inter-centroid-first}
    \end{subfigure}
    \hfill
    \begin{subfigure}{.49\linewidth}
        \centering
        \includegraphics[width=\linewidth]{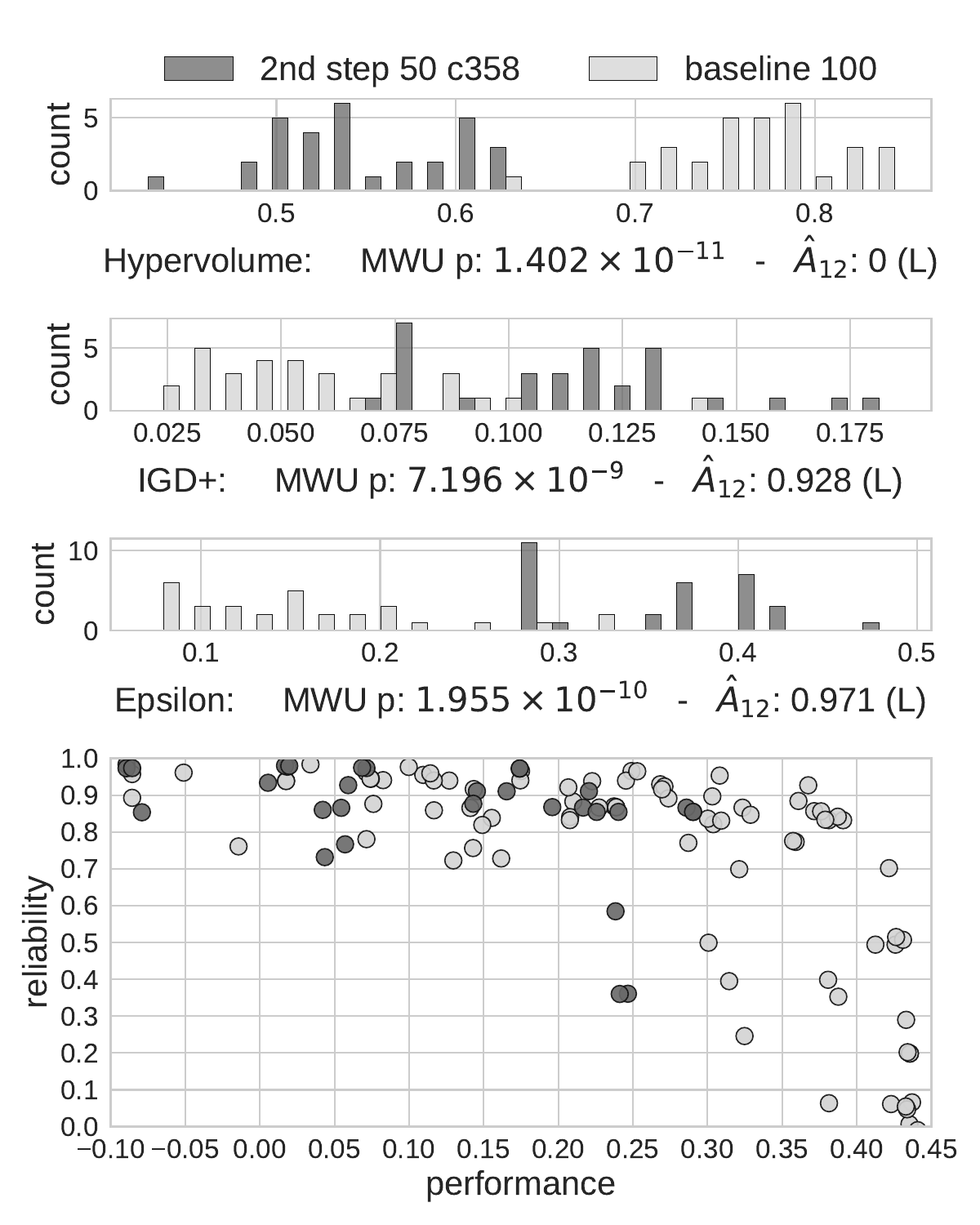}
        \caption{\ccsecondintcentrB against \base}
        \label{fig:rq1-ccm-inter-centroid-second}
    \end{subfigure}
    \caption{\ccm: comparison of the interactive steps based on selecting a cluster centroid against the \base, regarding \perfq, \reliability, and quality indicators. The upper half shows the distribution of the quality indicators among the (31) runs, along with the $p\-value$ of the \mwu test and the \vda effect size; magnitude interpretation: negligible (N), small (S), medium (M), large (L). The lower half reports the Pareto fronts obtained both from the interaction and from the \base, projected in a bidimensional space to compare the values of \perfq and \reliability.}
    \label{fig:rq1-ccm-interactive-centroid-vs-baseline}
\end{figure*}

This analysis is summarized by the plots in \Cref{fig:rq1-ttbs-interactive-centroid-vs-baseline,fig:rq1-ccm-interactive-centroid-vs-baseline}, which compare interactive experiments with \base.
On the upper half of these plots, we report the distribution of the quality indicator values among the 31 runs performed for each experimental setting, both for a single interactive experiment and for \base.
We applied the \mwu non-parametric statistical test \cite{Mann_Whitney_1947} where the null hypothesis states that the experiments do not have a statistically significant difference.
Two experiments are considered to be significantly different based on the values of a specific quality indicator when the test computes a p-value smaller than $\alpha=0.05$.
To assess the magnitude of the difference, we used the Vargha--Delaney \vda~\cite{Vargha_Delaney_2000}, a standardized non-parametric effect size measure.
\vda can take values between $0$ and $1$, and a value of $0.5$ indicates that the two experiments are equivalent.
The closer the value of \vda gets to $0$ or $1$, the larger the effect size.
The interpretation of \vda as negligible, small, medium, and large is performed according to the thresholds $0.147$, $0.33$, and $0.474$, respectively~\citep{Hess_Kromrey_2004}.

As it can be observed for \ttbs in \Cref{fig:rq1-ttbs-interactive-centroid-vs-baseline}, in all cases except for \igd in the \ttsecondintcentrB (\Cref{fig:rq1-ttbs-inter-centroid-second}), there is a significant difference between the interactive experiments and the \base with large effect size, despite some overlap of the distributions.
For \igd, such an overlap is more visible; for \ttsecondintcentrB, the distributions overlap almost entirely.
It can also be observed that the distribution yields a high variance, thus indicating a high variability in the explored space among the runs.
Moreover, regardless of the chosen centroid, the Pareto front is confined to two solutions in both interactive cases. One of the interactions (around $0.2$ \perfq and $0.7$ \reliability) has almost the same values of \perfq and \reliability, while the other has a smaller value of \perfq.
Interestingly, the two selected centroids led to excluding designs with low \reliability.
In both cases, the centroids had a relatively good value of \perfq, which was preserved (to some extent) in the very small front obtained from the interaction.
The variation in \perfq, when comparing the two interactive fronts, might be caused by the number of performance antipatterns associated with the centroids.
In fact, \ttsecondintcentrA started without (detected) performance antipatterns, while \ttsecondintcentrB had only two performance antipatterns, reported as \pas in \Cref{tab:cs_config}.
This most probably left the latter with more refactoring opportunities that effectively shrink the Pareto front toward better \perfq values.

The situation appears different for \ccm in \Cref{fig:rq1-ccm-interactive-centroid-vs-baseline}.
First, the distributions of the quality indicators show considerably less variance.
The \ccsecondintcentrA experiment (\Cref{fig:rq1-ccm-inter-centroid-first}) shows a mix of effect sizes, ranging from small to large.
On the contrary, \ccsecondintcentrB (\Cref{fig:rq1-ccm-inter-centroid-second}) always reports a large effect size.
In the first case, this means that depending on the specific quality indicator, we can be more or less sure of how distant the interactive front is from \base, whereas, in the second case, the results are more consistent.
This difference between the two interactive cases can also be found in the solution space.
The \ccsecondintcentrA interaction seems to have driven the search to the upper right corner toward better \perfq and \reliability values since almost all the front solutions are concentrated in that corner.
Conversely, in \ccsecondintcentrB the front is more spread, providing more dissimilarity among the alternatives.
However, in both cases, the fronts include very few solutions with a lower \reliability than \base.
The difference in the observed spread along the \perfq axis might be caused by the centroid selection during the interaction.
The centroids for the \ccsecondintcentrA and \ccsecondintcentrB experiments slightly differ in \reliability, but are on the opposite ends of the \perfq objective.
\ccsecondintcentrA had a very good starting value of \perfq, while \ccsecondintcentrB had among the lowest \perfq we observed.
This provided more room for \ccsecondintcentrB to spread along the \perfq axis while looking for better solutions, leading to the spread observed in the final Pareto front.

\begin{rqsummary}{Summary of RQ1}
As expected, the explored solution space is narrowed down by the designer's interactions.
Although the interactions might achieve less space coverage, in some cases (depending on the considered benchmark system) they drive the search to areas with more desirable solutions. 
The starting values of the designs before interaction obviously influence the areas of the solution space to be explored after the interaction, as well as the shape of the Pareto fronts. This influence occurs even in a fully automated approach, with the difference that multiple interactions in our approach make this influence heavier due to more frequent choices of re-starting points.
\end{rqsummary}
 \subsection{Architectural differences (RQ2)}\label{sed:results-rq3}

\begin{figure*}[htbp]
    \centering
    \begin{subfigure}{1.0\linewidth}
	\includegraphics[width=.98\textwidth]{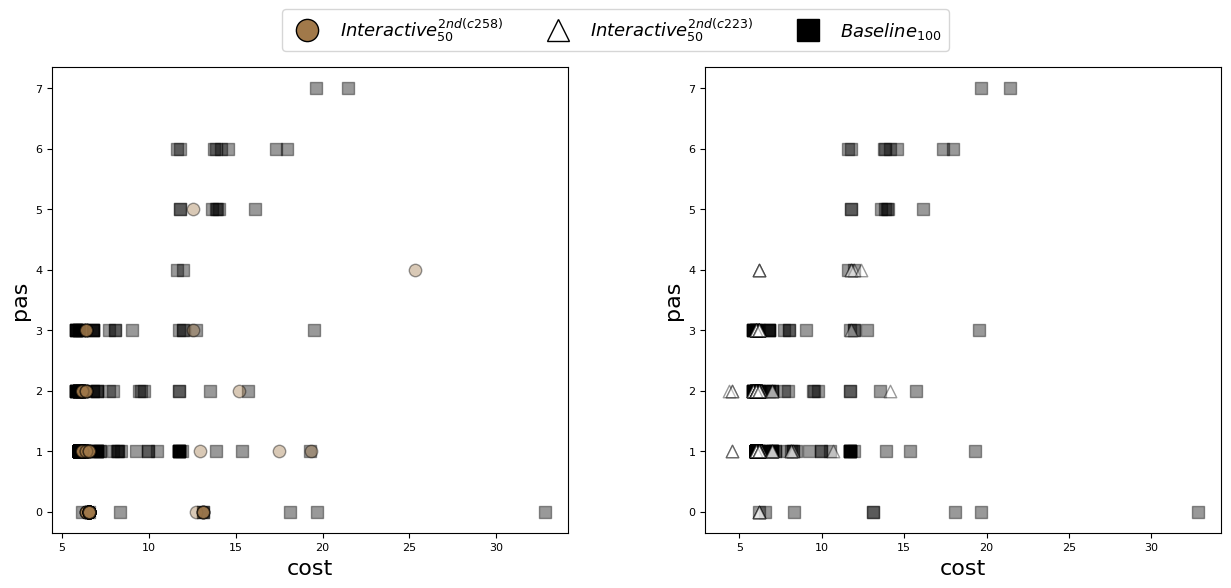}
	\caption{\ttbs\label{fig:scatter-trainticket}}
    \end{subfigure}\hfill
    \begin{subfigure}{1.0\linewidth}
	\includegraphics[width=.98\textwidth]{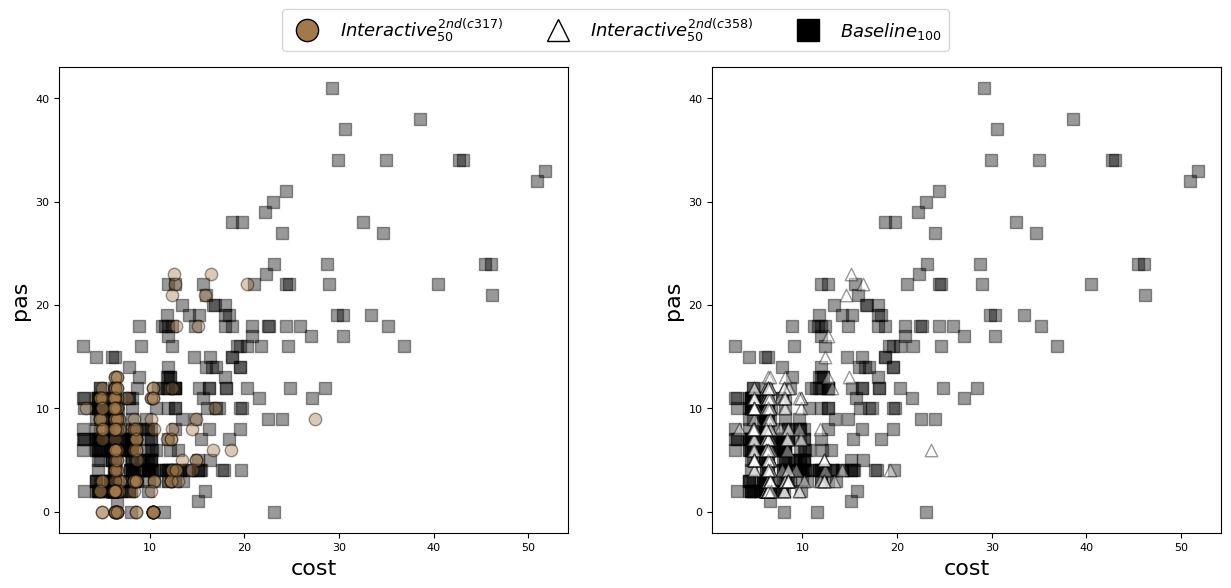}
	\caption{\ccm\label{fig:scatter-cocome}}
    \end{subfigure}
    \caption{Comparison of architectural models regarding \achanges and \pas characteristics. The brown circles (left sub-figure) and the white triangles (right sub-figure) correspond to the models derived from the choice of the two centroids, respectively, while the gray symbols correspond to \base.}
    \label{fig:scatters-changes-pas}
\end{figure*}

The scatter plots in \Cref{fig:scatters-changes-pas} show the solution spaces for the two benchmark systems when considering \achanges and \pas as high-level model characteristics. The symbols expose the differences between the \base (without interaction) and the interactive experiments. The latter covers a well-defined and narrower area (the left-bottom corner of the sub-figures) as compared to the area explored by the baseline, thus suggesting that centroid choices led to less variety in the explored designs. In particular, in \ccm these models are mostly confined to the ranges $[1.0,18.0]$ and $[0,20]$ for \achanges and \pas, respectively; while for \ttbs the models are mostly in the ranges $[3,12]$ and $[0,4]$, respectively. This is an initial indicator of space differences attributed to the designer's interaction, as reflected in \textit{RQ1}.

Based on the 5-point Likert scales for \achanges and \pas, we observed that the solutions resulting from the interactions focused on specific categories, while \base provided solutions covering several of such categories -- as hinted by the dispersion of the gray points in \Cref{fig:scatters-changes-pas}. We argue that the categories of the solutions were conditioned by (or correlated with) the characteristics of the selected centroids. In \ccm, both centroids had the same structural characteristics -- \textit{(very-few, very-few)} for \achanges and \pas, and the resulting solutions mainly fell under that category. A similar pattern was observed for the solutions derived from the centroids chosen for \ttbs. This means the models (after the interactions) are close to the initial architecture (regarding changes) and have very few antipatterns. Furthermore, the optimization algorithm that minimizes both architectural objectives drives the trend towards the \textit{(very-few, very-few)} category for \achanges and \pas.

\begin{figure}[htbp]
    \centering
	\begin{subfigure}{.48\textwidth}
	    \includegraphics[width=\textwidth]{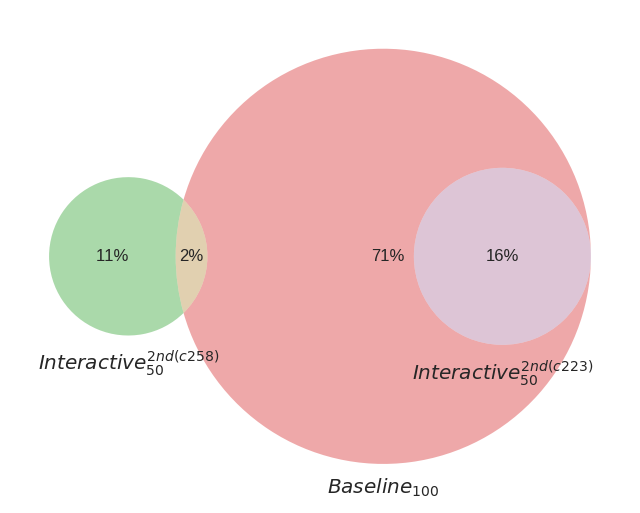}
	    \caption{\ttbs\label{fig:ttbs-intersections}}
	\end{subfigure}
	\hfill 
	\begin{subfigure}{.48\textwidth}
		\includegraphics[width=\textwidth]{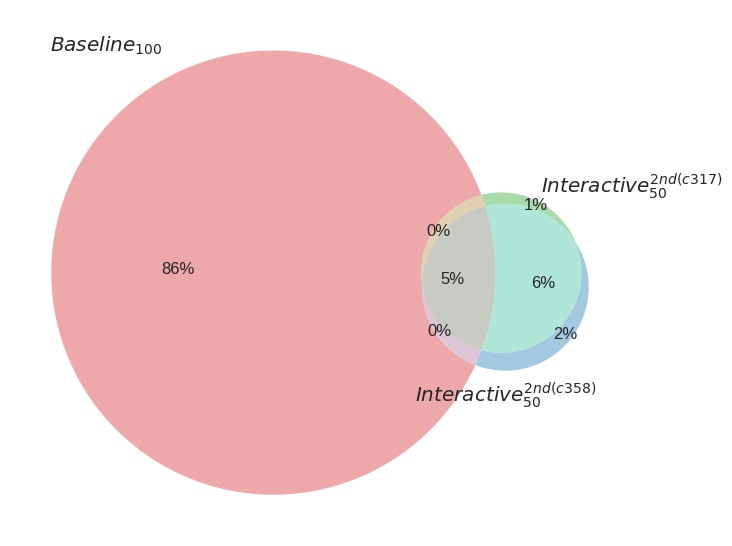}
		\caption{\ccm\label{fig:ccm-intersections}}
	\end{subfigure}
  
    \caption{Representation of the intersections between the sets of sequences of refactoring actions for different experiments. The large circle corresponds to \base (baseline), while the small circles correspond to the results after choosing the centroids. The circle sizes are proportional to the number of unique sequences per experiment.}
    \label{fig:refactions-intersections}
\end{figure}

The intersections among the sequences of refactoring actions from the different experiments are shown in the Venn diagrams of \Cref{fig:refactions-intersections}. In both benchmark systems, \base exhibited many more unique sequences than the sets resulting from the interactive experiments. The higher the number of unique sequences, the more diversity in the corresponding models. However, the interactive approach's lack of diversity in the designs explored is expected based on the experimental design. The form of interaction that we have set in this work (i.e., the preference-based interactivity) nudges the optimization to explore particular regions, and therefore, such lack of diversity is a feature, not a bug. Similar results have been reported in preference-based optimization literature~\citep{iqbal2020flexibo,abdolshah2019multi}.
In addition, we noticed that the two sets of sequences resulting from the \ccm interactive experiments were very similar between them (\Cref{fig:ccm-intersections}).

\Cref{tab:refactions-freq} reports the frequencies of the refactoring actions across the four positions of the sequences. Having a smaller number of unique sequences in the interactive experiments can be partial because the centroids ``freeze'' the two initial actions of the sequences. As underlined in \Cref{tab:refactions-freq}, actions \textit{MO2C} and \textit{MO2N} were selected for \ttbs (columns \textit{c-258} and \textit{c-223}), while \textit{MO2C} was selected for \ccm (columns \textit{c-317} and \textit{c-358}), respectively. 

Interestingly, the interactive experiments exhibited different frequency profiles than \base in both benchmark systems. The frequency variations are apparent in the third and fourth positions of the tables in \Cref{tab:refactions-freq}. For \ttbs, \textit{MO2N} was the most prevalent action in the three experiments, but it was less prevalent when picking the centroids in favor of \textit{MO2C}. The \base was exclusively focused on using \textit{MO2N}. For \ccm, there were variations in the frequency values, with \textit{MO2N} being the most prevalent action and some contributions of \textit{MO2C}. Along this line, we argue that the designer's interactions likely influence the composition of actions in the sequences.
The only exception was \ttsecondintcentrB in \ttbs (column \textit{c-223}), which has a similar profile as the baseline. This is correlated with \Cref{fig:ttbs-intersections}, in which the resulting sequences are all shared by the baseline.

\begin{table}[]
    \centering
    \caption{Frequencies of refactoring actions from Table \ref{tab:ref_actions} used at each position of the sequences for the two benchmark systems. The sequence position is
indicated by the \textit{pos} column. The frequencies are normalized. Each table compares the results of the two centroid choices against \base (baseline). Frequencies of 1.00 (maximum) are underlined.} \label{tab:refactions-freq}

\begin{tabular}{rlrrrcrlrrr}

\toprule

\multirow{2}{*}{\textbf{pos}} & \multirow{2}{*}{\textbf{ref}} & \multicolumn{3}{c}{TTBS} & & \multicolumn{3}{c}{CCM} \\
                                                                 \cline{3-5}                  \cline{7-9}
	&              & \textbf{baseline} & \textbf{c258} & \textbf{c223}  & & \textbf{baseline} & \textbf{c317} & \textbf{c358}     \\ 

\midrule
\multirow{4}{*}{1}  & ReDe & 0.03 & 0.00             & 0.00              & & 0.21  & 0.00             & 0.00             \\ 
		    & MO2C & 0.10 & \underline{1.00} & 0.00              & & 0.12  & \underline{1.00} & \underline{1.00} \\ 
		    & MO2N & 0.84 & 0.00             & \underline{1.00}  & & 0.64  & 0.00             & 0.00             \\ 
                    & Clon & 0.03 & 0.00             & 0.00              & & 0.03  & 0.00             & 0.00             \\ 
\midrule                                                                                                                 
\multirow{4}{*}{2}  & ReDe & 0.03 & 0.00             & 0.00              & & 0.16  & 0.00             & 0.00             \\ 
		    & MO2C & 0.07 & \underline{1.00} & 0.00              & & 0.25  & \underline{1.00} & \underline{1.00} \\
		    & MO2N & 0.87 & 0.00             & \underline{1.00}  & & 0.56  & 0.00             & 0.00             \\
                    & Clon & 0.03 & 0.00             & 0.00              & & 0.03  & 0.00             & 0.00             \\ 
\midrule                                                                                                                 
\multirow{4}{*}{3}  & ReDe & 0.04 & 0.01             & 0.03              & & 0.22  & 0.28             & 0.19             \\ 
                    & MO2C & 0.06 & 0.36             & 0.45              & & 0.23  & 0.25             & 0.37             \\ 
                    & MO2N & 0.85 & 0.63             & 0.51              & & 0.53  & 0.46             & 0.43             \\ 
                    & Clon & 0.04 & 0.00             & 0.01              & & 0.02  & 0.01             & 0.01             \\ 
\midrule                                                                                                                 
\multirow{4}{*}{4}  & ReDe & 0.02 & 0.00             & 0.02              & & 0.21  & 0.31             & 0.19             \\ 
                    & MO2C & 0.33 & 0.33             & 0.51              & & 0.45  & 0.22             & 0.35             \\ 
                    & MO2N & 0.64 & 0.66             & 0.46              & & 0.32  & 0.47             & 0.45             \\ 
                    & Clon & 0.02 & 0.00             & 0.01              & & 0.01  & 0.00             & 0.01             \\ 
\bottomrule
\end{tabular}
\end{table}

\begin{rqsummary}{Summary of RQ2}
The \achanges and \pas characteristics of the architectural models generated by the designer's interactions are a subset of the range of characteristics explored by the baseline. These characteristics are the categories of values resulting from 5-point scales. In particular, the subset refers to very low costs and very few performance antipatterns. At a fine-grained level, the architectural models, as represented by their underlying sequences of refactoring actions, also present structural differences concerning the baseline. Although the centroids freeze the first actions of the sequences, they enable variations in the remaining actions. As a result, we observed new sequences (and potentially, new underlying models) derived from the designer's interactions. \end{rqsummary}

 \subsection{Coverage of solution space (RQ3)}\label{sec:results-rq4}

\begin{figure*}[htbp]
    \centering
	\begin{subfigure}{.33\textwidth}
	    \includegraphics[width=0.9\textwidth]{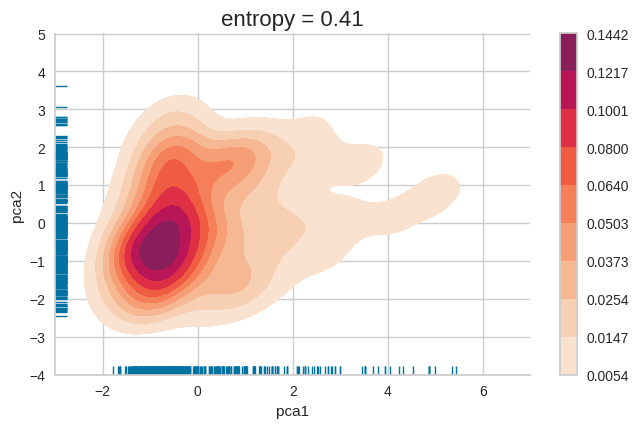}
	    \caption{\base \label{fig:density-baseline-cocome}}
	\end{subfigure}~
	\begin{subfigure}{.33\textwidth}
		\includegraphics[width=0.9\textwidth]{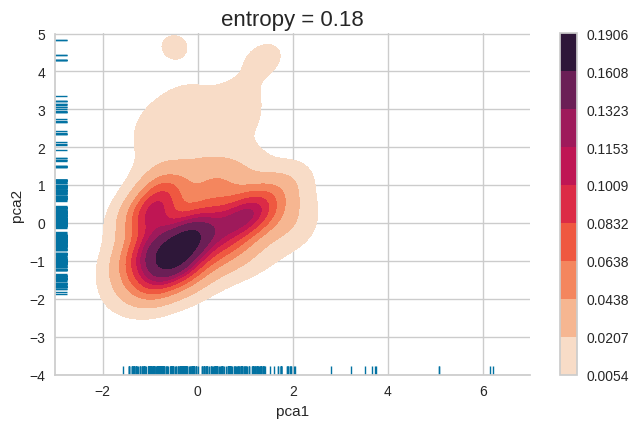}
		\caption{\ccsecondintcentrB \label{fig:density-c358-cocome}}
	\end{subfigure}~
        \begin{subfigure}{.33\textwidth}
		\includegraphics[width=0.9\textwidth]{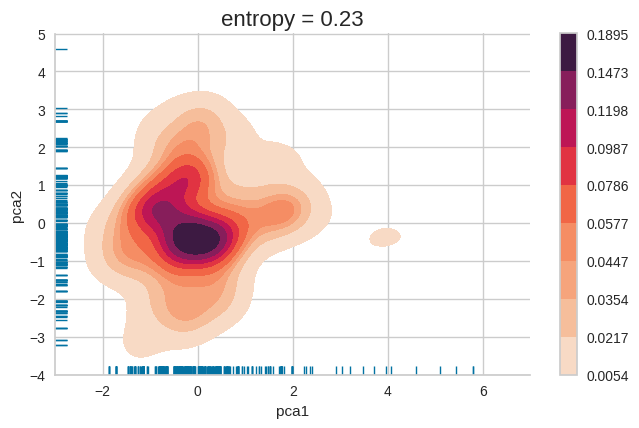}
		\caption{\ccsecondintcentrA \label{fig:density-c317-cocome}}
	\end{subfigure}\hfill
 
	\caption{Visualization of the objective landscapes for \ccm using KDE plots, after a 2D PCA projection (including standardization) of the original data. The heatmap reflects the concentration of points in certain regions. The blue rugs on each axis capture the distribution of the points.}
    \label{fig:landscape-cocome}
\end{figure*}

\begin{figure*}[htbp]
    \centering
	\begin{subfigure}{.33\textwidth}
	    \includegraphics[width=0.9\textwidth]{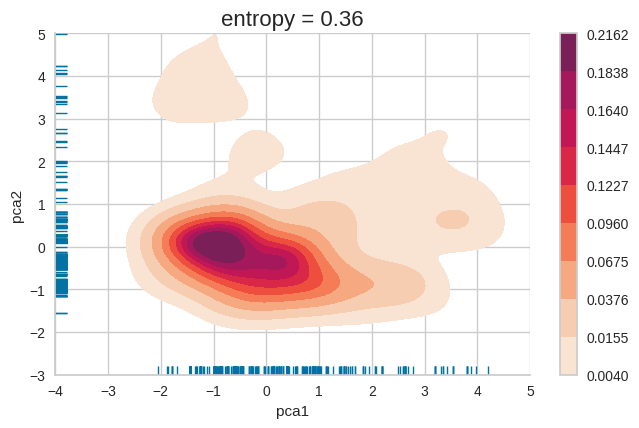}
	    \caption{\base \label{fig:density-baseline-trainticket}}
	\end{subfigure}~
	\begin{subfigure}{.33\textwidth}
		\includegraphics[width=0.9\textwidth]{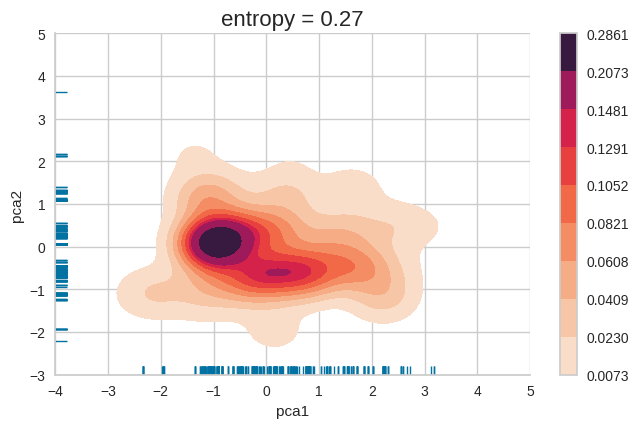}
		\caption{\ttsecondintcentrB \label{fig:density-c223-trainticket}}
	\end{subfigure}~
        \begin{subfigure}{.33\textwidth}
		\includegraphics[width=0.9\textwidth]{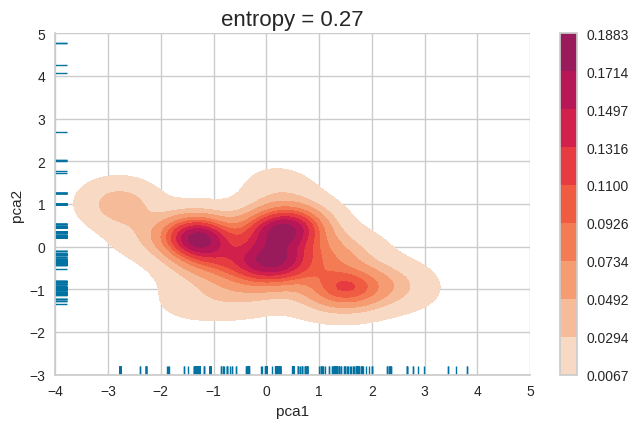}
		\caption{\ttsecondintcentrA \label{fig:density-c258-trainticket}}
	\end{subfigure}\hfill
 
    \caption{Visualization of the objective landscapes for \ttbs using KDE plots, after a 2D PCA projection (including standardization) of the original data. The heatmap reflects the concentration of points in certain regions. The blue rugs on each axis capture the distribution of the points.}
    \label{fig:landscape-trainticket}
\end{figure*}

The distributions of the objective values over the solution spaces of both benchmark systems are depicted, as KDE heatmaps, in \Cref{fig:landscape-cocome} and \Cref{fig:landscape-trainticket}. In these PCA projections, the variance captured by the first two principal components (i.e., the axes in the charts) was $89\%$ and $76\%$ for \ccm and \ttbs, respectively. Note that the resulting spaces have been all standardized for comparison purposes. A visual analysis reveals differences in (the shapes and colors of) the landscapes. \base spans a larger area than the remaining landscapes generated by the interactive experiments. The concentration of the solutions, which defines darker regions in heatmaps, also exhibits different patterns. While \base reveals a single and extended concentration area, the landscapes derived from the centroids show several smaller concentration areas. These patterns indicate that the centroid choice intensifies the search in certain regions, supported by the rugs' span on the different charts' axes. This, intuitively, indicates that interaction resulted in exploring multiple areas of interest that have optimal solutions, but they differ considerably in terms of the sequence of refactoring actions that were applied to reach the solution. 

The entropy values seem to corroborate the observations above. In \ccm, a value of $0.41$ for the \base suggests a homogenous trade-off distribution and the lack of very high concentrations of solutions (or peaks). In \ccsecondintcentrA and \ccsecondintcentrB, the entropy values decrease while the space coverage decreases and some peaks emerge. \base covers around 38 different trade-offs (out of 365) for the four objectives, being \textit{(slow, very-reliable, very-few, very-few)} the most prevalent one ($38\%$). When it comes to the interactions, \ccsecondintcentrA covers only 13 trade-offs with \textit{(slow, very-reliable, very-few, very-few)} and \textit{(average, very-reliable, very-few, very-few)} as the most representative categories ($42\%$ and $32\%$, respectively). In \ccsecondintcentrB, the main trade-off is \textit{(slow, very-reliable, very-few, very-few)} and accounts for $73\%$ of the solutions.

In \ttbs, the space coverage is slightly different among the three experiments, but the concentration of solutions and the peaks in the landscapes are more similar. Thus, the entropy values are closer across experiments. Nonetheless, we can still notice some intensification regions resulting from the interactions in the landscapes. 

\begin{figure*}[htbp]
    \centering
	\begin{subfigure}{.33\textwidth}
	    \includegraphics[width=0.9\textwidth]{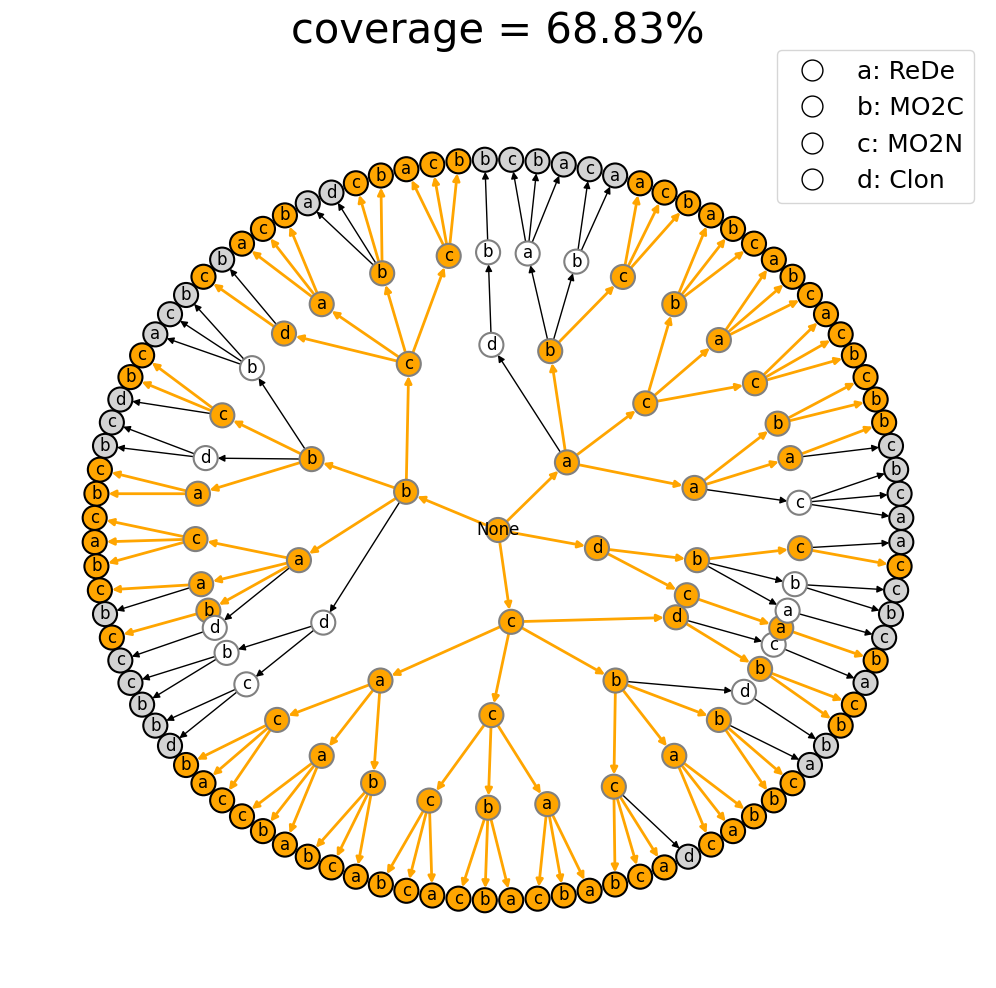}
	    \caption{\base \label{fig:trie-baseline-cocome}}
	\end{subfigure}~
	\begin{subfigure}{.33\textwidth}
		\includegraphics[width=0.9\textwidth]{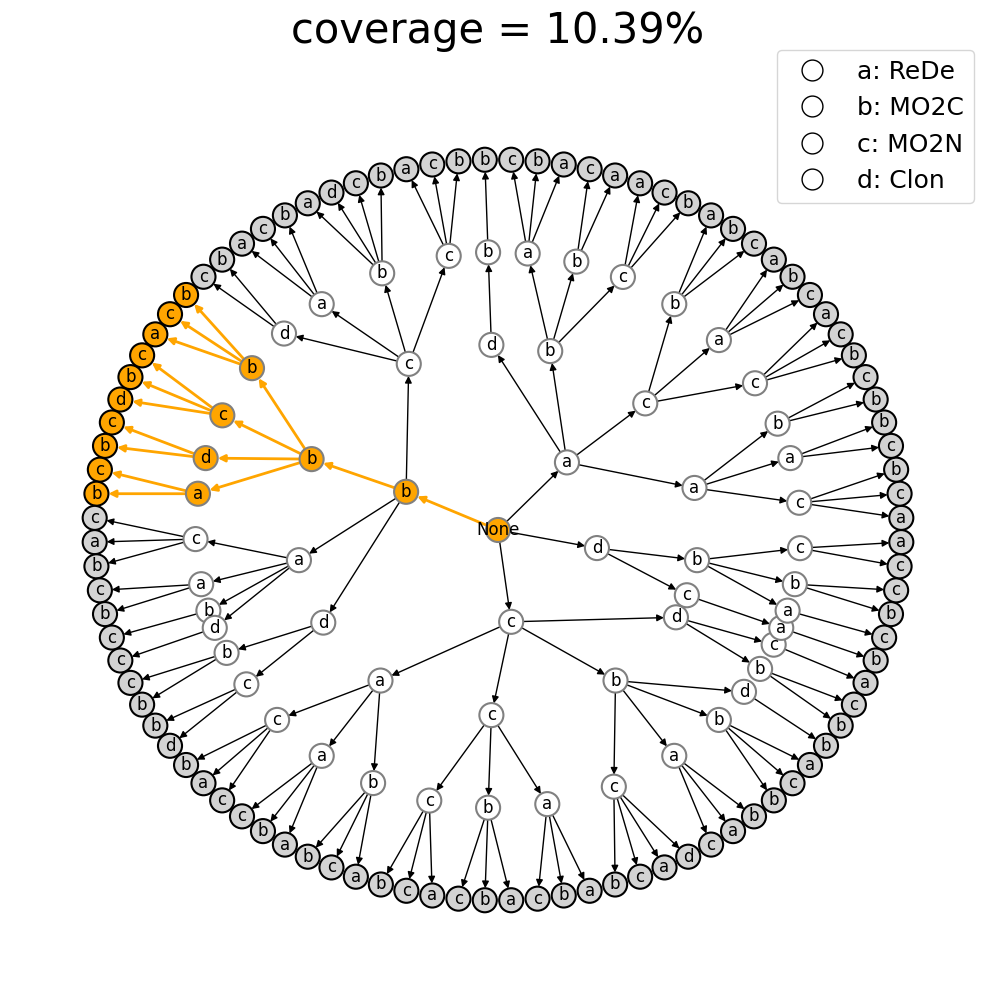}
		\caption{\ccsecondintcentrB \label{fig:trie-c358-cocome}}
	\end{subfigure}~
        \begin{subfigure}{.33\textwidth}
		\includegraphics[width=0.9\textwidth]{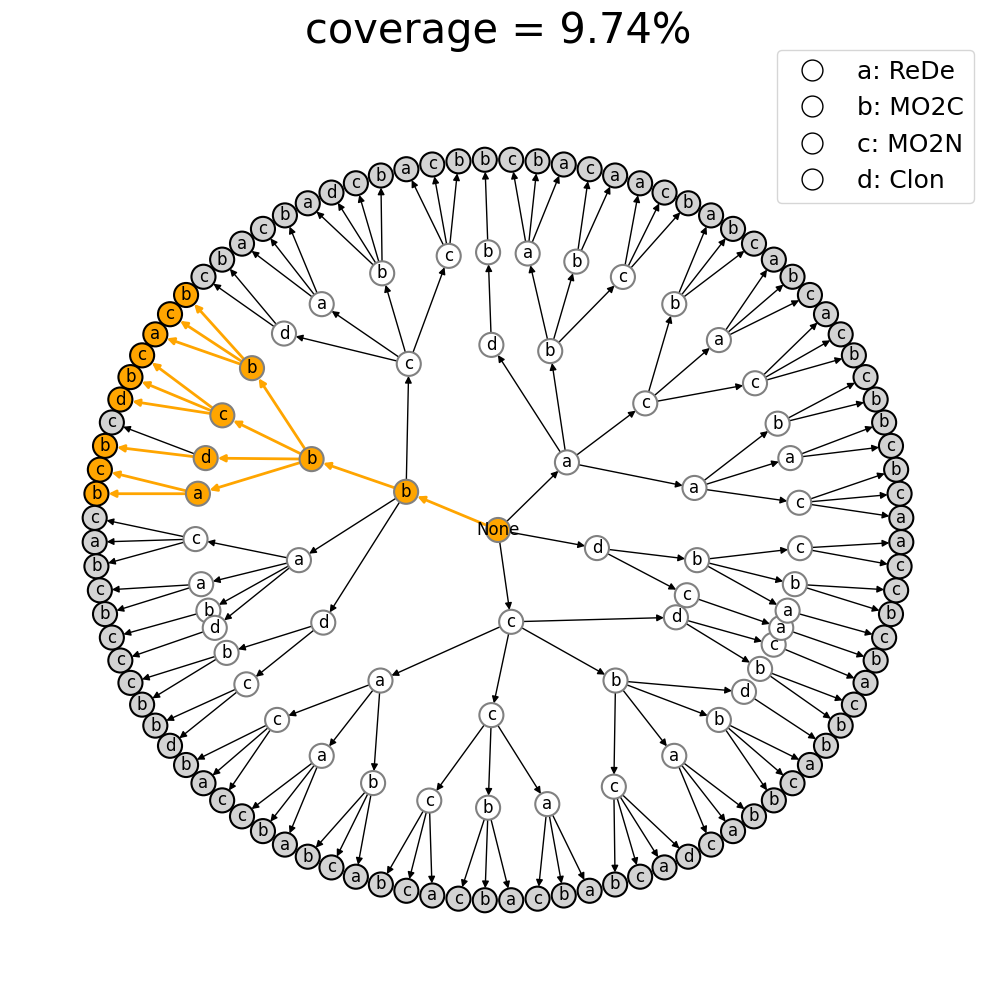}
		\caption{\ccsecondintcentrA \label{fig:trie-c317-cocome}}
	\end{subfigure}\hfill
 
	\caption{Sequences of refactoring actions for \ccm represented as trees (as generated by \nsga). Each node maps to an individual refactoring action, as indicated in the legend. The node size is proportional to the number of architectural models resulting from that particular path (i.e., a sequence). The orange nodes and edges correspond to the sequences of refactoring actions that intersect with those in \refpf. The larger the intersection, the higher the coverage of the space.}
    \label{fig:search-trees-cocome}
\end{figure*}

\begin{figure*}[htbp]
    \centering
	\begin{subfigure}{.33\textwidth}
	    \includegraphics[width=0.9\textwidth]{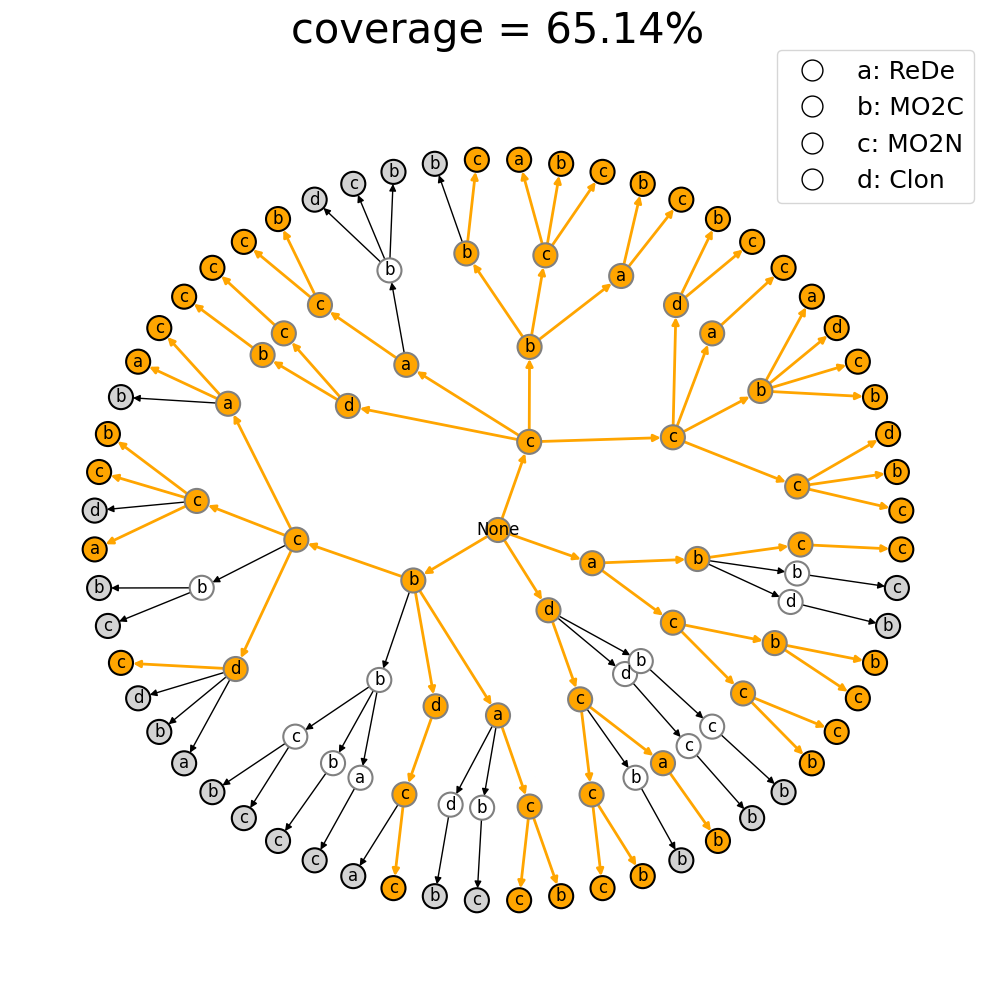}
	    \caption{\base \label{fig:trie-baseline-trainticket}}
	\end{subfigure}~
	\begin{subfigure}{.33\textwidth}
		\includegraphics[width=0.9\textwidth]{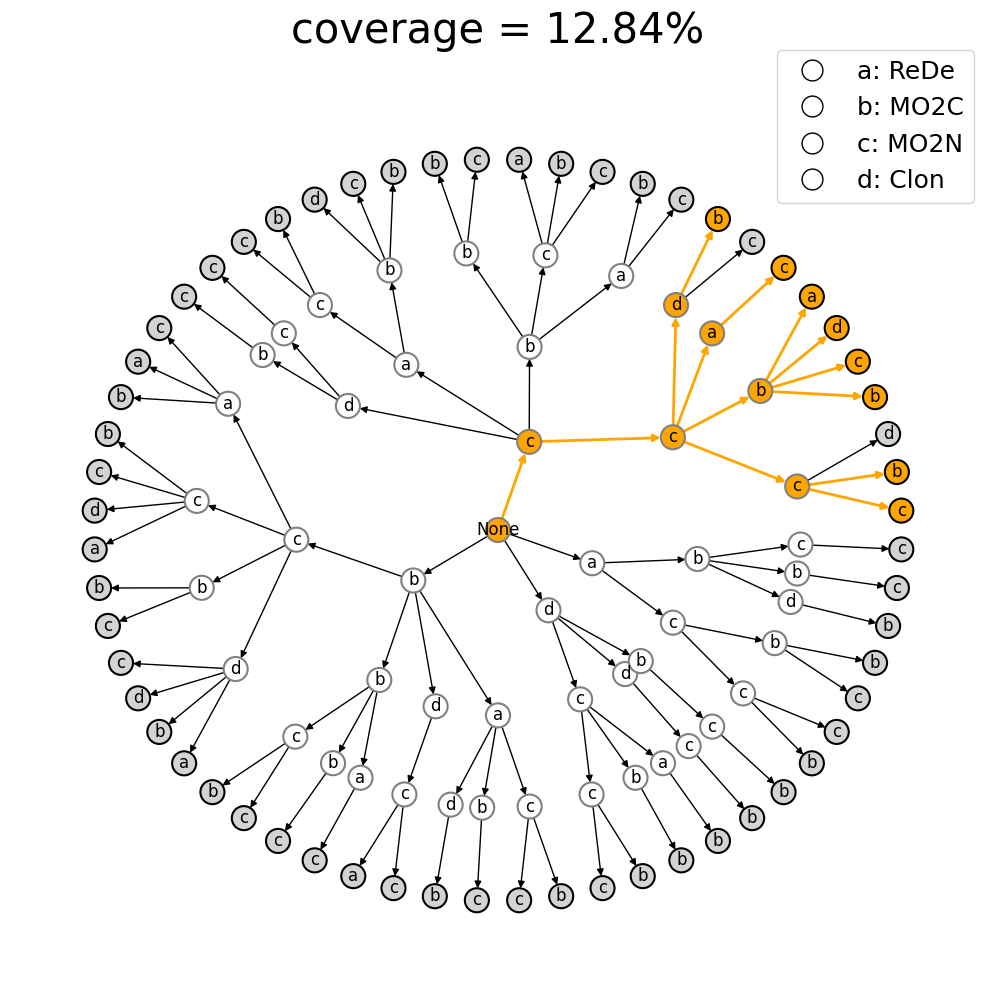}
		\caption{\ttsecondintcentrB \label{fig:trie-c223-trainticket}}
	\end{subfigure}~
        \begin{subfigure}{.33\textwidth}
		\includegraphics[width=0.9\textwidth]{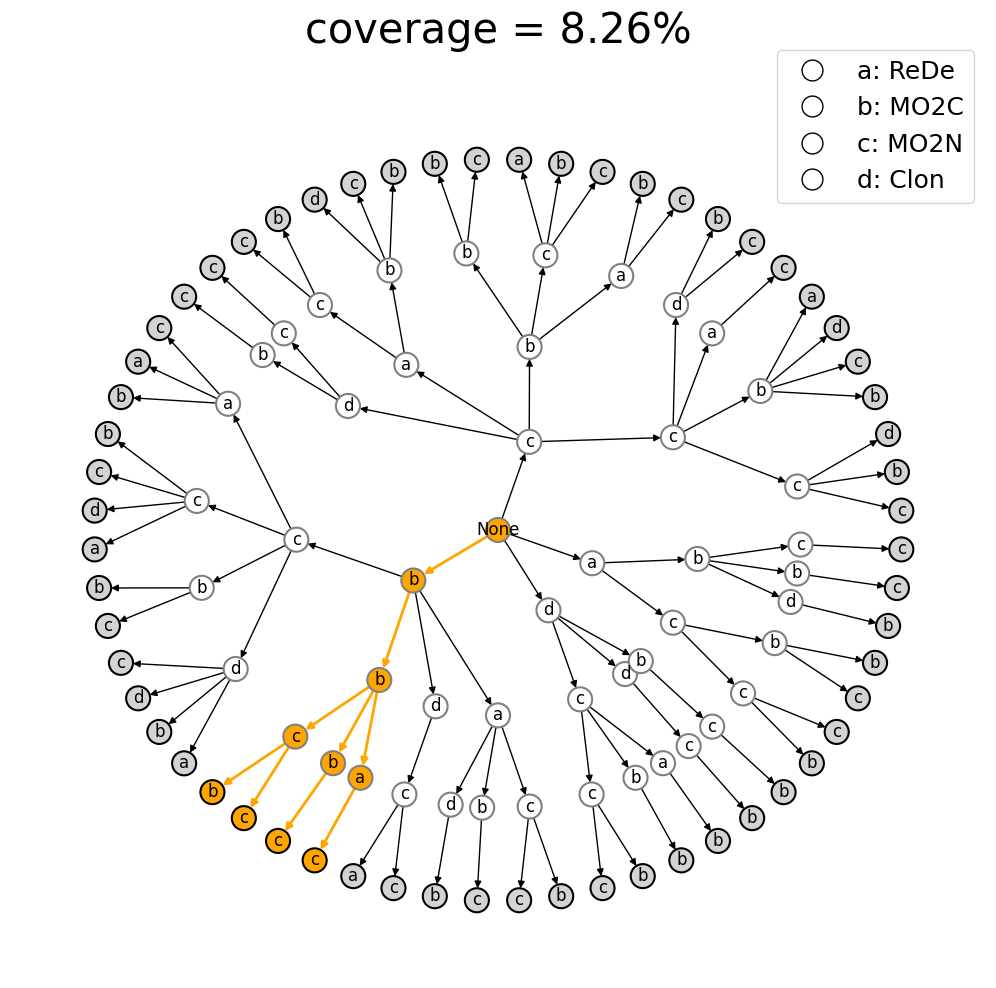}
		\caption{\ttsecondintcentrA \label{fig:trie-c258-trainticket}}
	\end{subfigure}\hfill
 
    \caption{Sequences of refactoring actions for \ttbs represented as trees (as generated by \nsga). Each node maps to an individual refactoring action, as indicated in the legend. The node size is proportional to the number of architectural models resulting from that particular path (i.e., a sequence). The orange nodes and edges correspond to the sequences of refactoring actions that intersect with those in \refpf. The larger the intersection, the higher the coverage of the space.}
    \label{fig:search-trees-trainticket}
\end{figure*}

 \Cref{fig:search-trees-cocome,fig:search-trees-trainticket} present the reference tree for each system, showing the coverage levels achieved by the experiments without and with interactions. As expected and confirmed graphically, the sequences after choosing the centroids are focused on very small subtrees, while \base covers more than two-thirds of the reference space. The figures for coverage follow the same trend in both benchmark systems. In \ccm, the subtree explored by both \ccsecondintcentrA and \ccsecondintcentrB has only a small overlap with the nodes covered by \base, which indicates that new models are returned due to the user interaction. A similar situation happens for \ttbs with the subtree explored by \ttsecondintcentrA.
From \Cref{tab:coverage}, the coverage metric shows that in \ttsecondintcentrA (for \ttbs) the new models found by the designer's interactions dominate part of the Pareto front of \base.
These observations suggest that the designer's choices steered the search toward models that would not be found without interactions.
However, we also recognize that, in most cases, the \base covers a larger portion of the space than the interactive ones. This is expected, as the \base explores the space without any constraints, while interactions focus on specific areas.
Furthermore, we should note that the computational cost for exploring these ``new'' models is relatively low when compared to the costs incurred for \refpf, which requires an extensive search in the space. In other words, the designer's interaction worked as a mechanism to spend the computational costs in the exploration of preferred types of solutions (as marked by the centroids).

\begin{table}[htbp]

    \newcommand{\cov}[2]{{\footnotesize $C($#1, #2$)$}}
    \centering
    \caption{Values of the coverage metric for interactions against the baseline.}
    \label{tab:coverage}
    \begin{tabular}{lcc}
        \toprule
	\multirow{4}{*}{\textbf{CCM}}  & \cov{\ccsecondintcentrA}{\base} & \cov{\base}{\ccsecondintcentrA} \\
                      & $0$                          & $0.923077$ \\
                      \cline{2-3} \noalign{\vskip 3pt}
                      & \cov{\ccsecondintcentrB}{\base} & \cov{\base}{\ccsecondintcentrB} \\
                      & $0.011236$                   & $0.733333$ \\
        \midrule
	\multirow{4}{*}{\textbf{TTBS}} & \cov{\ttsecondintcentrA}{\base} & \cov{\base}{\ttsecondintcentrA} \\
                      & $0.214286$                   & $0$ \\
                      \cline{2-3} \noalign{\vskip 3pt}
                      & \cov{\ttsecondintcentrB}{\base} & \cov{\base}{\ttsecondintcentrB} \\
                      & $0$                          & $0.5$ \\
        \bottomrule
    \end{tabular}

\end{table}

As a final consideration we observe that, although the improved efficiency of an interactive process is not a new insight per se, it was not yet illustrated in a software architecture optimization context.

Beside this, there is a spectrum of possibilities between pre-defined and post-defined preferences. The former case, where designer's preferences are defined before the optimization process and cannot be modified, can result more efficient in terms of processing time, but it suffers from stiffness in terms of priorities/weights of different objectives. As opposite, an interactive approach obviously introduces time overhead to the whole process, but it enables flexibility in the designer's preferences that can change while the process runs. 

This knob in the hands of designers would realistically produce more benefits when the objective space cannot be analytically expressed due to its intrinsic complexity. As an example in software architecture optimization, this is the case of performance indices that typically emerge from the analysis of complex models (e.g., Queueing Networks). In these cases, an interactive approach enables designers to manage this emerging complexity while the optimization progresses, thus steering the search towards desirable directions.

\begin{rqsummary}{Summary of RQ3}
The space of architectural models and objective values explored without the designer's interactions have a larger coverage due to the spread of the data points than the space resulting from the interactions. The interactions intensify the search in predetermined space areas, thus leading to (more) solutions with specific trade-offs. Despite having a tiny coverage, the interactions induced the searching process to reach architectural models not seen in the baseline, which becomes a desirable output for the designer. 
\end{rqsummary}

 \section{User study}\label{sec:results-userstudy}
\begin{figure}[htbp]
    \centering
    \includegraphics[width=\linewidth]{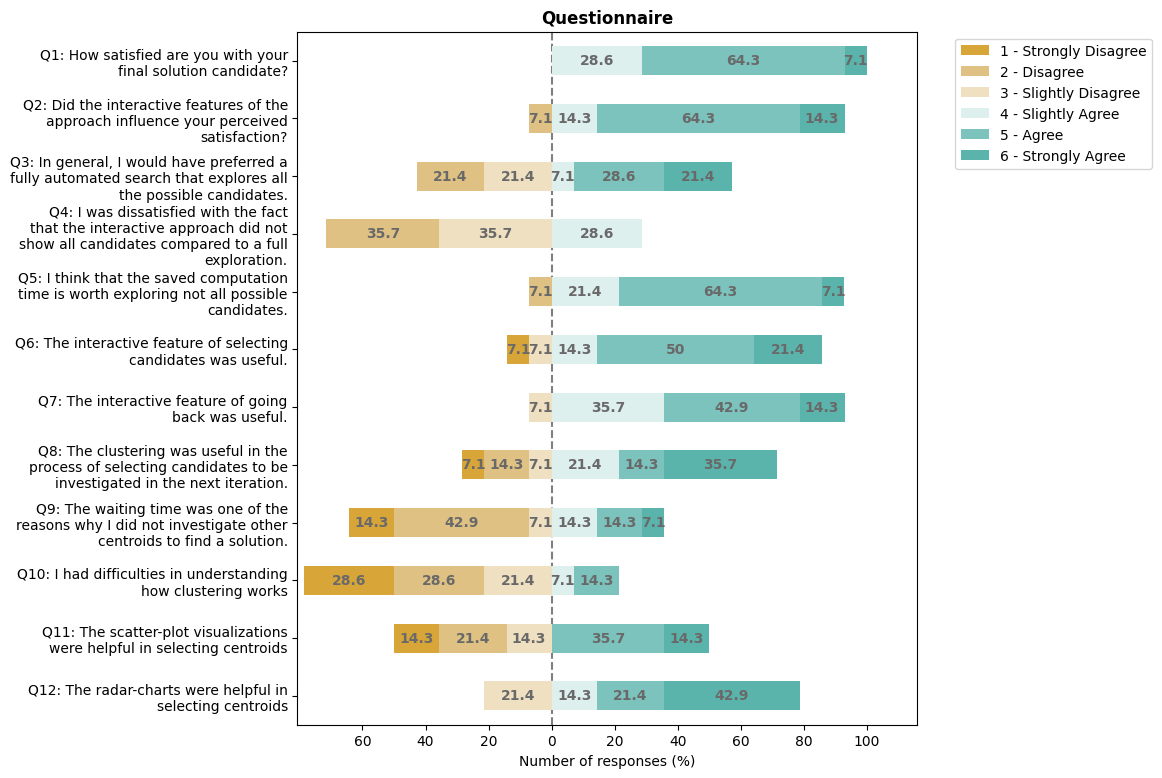}
    \caption{Results from the questions asked after the exploration session (user study).}\label{fig:results-userstudy}
\end{figure}

The results of the participants' answers to the questionnaire are displayed in Figure \ref{fig:results-userstudy}. From questions \textit{Q1} and \textit{Q2} we can observe that participants were satisfied with the solution candidates explored during the design sessions, and they were satisfied with the interaction mechanisms provided by the \textit{Jupyter} notebook. 

When it comes to the comparison to a fully-automated approach, in questions \textit{Q3}, \textit{Q4} and \textit{Q5}, participants' opinions were divided. It seemed participants were not affected by the fact that only some of the possible solutions were shown by the visualization charts. Almost half of the participants agreed with the argument that the interactions should save a considerable computational effort to be useful during the exploration. This fact, along with question \textit{Q6}, supports the interactive mechanism proposed by our approach, although we cannot determine how much effort can be actually saved. Although participants had to wait (for a short period of time) whenever they explored a new branch of the tree, this was not perceived as an obstacle, as indicated by question \textit{Q9}. Based on participants' feedback, we noticed that they were driven by particular preferences, and some of those preferences even changed as long as choices were made; thus, we conjecture that participants were willing to wait for the optimization at certain nodes as long as they perceived that unveiling solutions under that node was useful.

During the sessions, we had to make clarifications about how the clustering of solutions worked and about the role of centroids, as some participants experienced issues with the usage of the technique. Overall, question \textit{Q10} indicates that clustering was not a major obstacle for the exercise. The positive effect of clustering was assessed by question \textit{Q8}. Along this line, in question \textit{Q12}, the radar charts were reported as helpful for navigating the tree. 

Furthermore, an observation from question \textit{Q7} was about the need for complemeting the interactive mechanism with navigation-related features, such as the ability to go back in the tree offered by the notebook. On the downside, as indicated by \textit{Q11}, the information conveyed by the scatter plots was difficult to grasp for some participants, probably due to information overloading issues. These comments suggest areas of improvement for the user experience of the designer, in addition to providing interactive mechanisms in the optimization process.

\begin{rqsummary}{Summary of RQ4 (user study)}
The designers were generally satisfied with the solution candidates and the interactive process. The study participants accepted our interactive process' trade-off between sufficient exploration of the solution space and the required computation time. Regarding the provided interactive features, most participants found selecting candidates and reverting the selection useful for a better exploration of the solution space. Most of the participants appreciated the clustering of solutions as a way of simplifying the search space. \PJ{appreciated in doing what?}
\end{rqsummary}
  \section{Threats to Validity}\label{sec:threats}

We classify threats that might affect our approach, as described by \citet{DBLP:books/daglib/0029933}.

\paragraph*{Construct validity} The parameters we used to augment the initial models with non-functional information might have affected our results, since all the objectives are computed by considering such information. Model parametrization from the literature (\eg \citep{Di_Pompeo_Tucci_2022}) was used to mitigate this threat.

Construct validity may also pertain to the ability of our proposed approach to explore the trade-offs between different quality attributes.
Specifically, the choice of the genetic algorithm used in the automated parts of the approach might impact how solutions are generated by \easier, as well as the space explored during the search.
We employed \nsga, which is among the most used algorithms~\citep{Ouni_Kessentini_Cinneide_Sahraoui_Deb_Inoue_2017,Koziolek_Koziolek_Reussner_2011}, and it has proved to perform well in similar studies~\citep{Corne_Jerram_Knowles_Oates_2001,Hiroyasu_Nakayama_Miki_2005}.
Furthermore, we followed best practices from the literature to reduce the impact of the algorithm on the final results~\citep{Arcuri_Fraser_2011,Arcuri_Briand_2011}.
In our approach, NSGA-II represents \emph{just an instance} of a possible search algorithm that can be employed to explore the solution space. In other words, the algorithm of choice is \emph{pluggable} within the approach, without the need to modify anything else. All we need is for the selected algorithm to be able to search the solution space and yield a Pareto front. Therefore, in this work, while we report its performance, we do not claim that NSGA-II is the best algorithm for the task.
In fact, recent studies have shown that NSGA-II might not always show the best performance, especially when dealing with many-objective optimization problems~\citep{PangIS20a}.
We leave the exploration of other algorithms for future work.

The usage of sequences of refactoring actions (in \textit{RQ2} and \textit{RQ3}) as proxies for the architectural models being explored constitutes a threat to construct validity. Furthermore, we only analyzed the types of refactoring actions (in \textit{RQ2} and \textit{RQ3}), and disregarded the parameters (or targets) of those actions. Because of this representation, certain sequences might enclose several architectural models that are not necessarily alike in terms of their structure. It could also be the case that two different sequences lead to the same UML model. Considering a finer-grained representation of the refactoring actions and including their logical dependencies can help to alleviate this threat.

Another potential threat to construct validity arises from the high correlation between the \textit{cost} and \textit{pas} objectives in the \ccm application (Pearson's correlation coefficient of $0.8377$). Since these objectives are strongly correlated, they may not represent distinct optimization goals, potentially leading to a less diverse set of solutions. This could hinder the algorithm's ability to explore meaningful trade-offs between objectives. A different formulation of the objectives might reduce this redundancy and yield better solutions. However, it is important to note that this issue does not occur in the \ttbs application, where the objectives are more independent (Pearson's correlation coefficient of $0.096$). Since we aim to provide a general, application-agnostic formulation, we chose not to reformulate the objectives for specific cases like \ccm, as doing so would undermine the broad applicability of our approach.

\paragraph*{Internal validity}

The fact that the sequences of refactoring actions had a fixed length (i.e., two or four actions per sequence) facilitated the comparisons, but it might have conditioned the results of the experiments. Allowing sequences with varying and longer lengths could lead to other architectural models and solutions in the explored spaces.

When it comes to \textit{RQ3}, a caveat of the entropy metric is the symmetry of the entropy definition, whose computation does not discriminate among landscapes with similar features but in different space locations. Thus, the interpretation of the entropy results should take into account that some spatial distributions of quality-attribute trade-offs might have gone undetected in our analysis.

\paragraph*{External validity}

We considered a relatively small catalog of refactoring actions. 
While having four refactoring actions might not seem enough to capture complex modifications of the initial architectural model, the size of the solution spaces (with respect to the number of possible refactoring sequences to apply) already represents a challenge in terms of time and computational costs~\citep{Guerrero_Marti_Berlanga_Garcia_Molina_2010,Marti_Garcia_Berlanga_Molina_2009,Wagner_Trautmann_Naujoks_2009}.
In future work, we plan to introduce other refactoring actions in our portfolio to mitigate this threat.

The number of benchmark systems and the size of the refactoring portfolio might reduce the validity of our findings.
We validated our approach on two benchmark systems only.
While these benchmark systems were selected from the literature and differ in size and complexity, they might not represent some additional challenges our approach could face in practice. We should use additional systems to mitigate this threat, which might also strengthen the generalization of our results.

In the user study, the fact that the participants had little industry experience can point the findings of the study towards a more academic architectural domain, limiting their generalizability. Furthermore, the experiments were based only on one system (\ttbs). We expect to replicate the study with a group of practitioners to assess possible differences in their perception of the interactions supported by our approach. We simplified parts of the tool setup, such as the clustering and navigation mechanisms and the usage of pre-computed solutions, to alleviate the participants' learning curve and facilitate their work and outcomes. In doing so, participants might not have experienced the full computational and cognitive efforts of a multi-objective optimization process.

\paragraph*{Conclusion validity}
There is a threat related to the interaction simulated in this study, as we did not involve real users in the interactive optimization process nor in the assessment of the resulting architectural models (e.g., with respect to their component structure). To mitigate this threat, two authors performed the interaction to simulate an engineering process aimed at improving the performance and reliability of a given software architecture. The architectural models, in turn, were analyzed in terms of the refactoring actions that generated them.

In addition, the choice of the cluster centroids was the only interaction mode considered in the experiments, and two (from the four available) centroids were considered. Other interaction modes (\eg dropping a given objective) could have been exercised for the comparisons \citep{Ramirez_Romero_Simons_2019}. This alternative should be explored in future work.

Another threat is related to the estimation of the quality of the Pareto fronts. 
To deal with this threat, we employed well-known quality indicators in \textit{RQ1}. We computed them assuming that the best Pareto fronts observed in long experiments are reasonably close to the optimal fronts for our benchmark systems.
While this might not always be the case, our conclusions were drawn by comparing the relative values of such indicators, relying on how close an experiment was to the best solutions we ever obtained.

  \section{Conclusion}\label{sec:conclusion}

In this work, we investigate whether designers can influence the output of an architecture optimization process through a preference-based interaction. To do so, we took two existing benchmark systems from the literature and performed several experiments, where we analyzed the differences between a fully automated process and an interactive one. We materialized the interactions as the choice of representative solutions (cluster centroids) from the space. We observed correlations between the interactions and the characteristics of the resulting solutions compared to the solutions generated by a baseline optimization process (without interactions). According to our experiments, the interactions seem to drive the search towards narrow areas of the objective space, which have less variability in their ranges of values. As the search intensifies in those areas, the interactions can contribute to the discovery of new architectural models that might lead to new trade-offs regarding the objectives. Thus, a key research finding is that steering the optimization by anchoring specific solutions (for subsequent optimization rounds) can be beneficial in terms of solution diversity, but it can also reduce the computational costs spent in the exploration. The software architects that participated in our user study agreed that this reduction in computational costs is worth refraining from a full exploration of the solution space, and they considered advantageous our employed interactive features. This indicates the usefulness of conducting interactive design processes in practice.

A lesson learned from this work is about the challenge of assessing the quality of the solutions. Although there are several quality indicators for the objectives in the literature (as used in \textit{RQ1}), less attention has been paid to indicators of the quality of architectural models resulting from the optimization process. For instance, \citet{CortellessaDPT:ecsa2023} proposed a quality indicator for the spread, namely the maximum architectural spread (MAS), that assesses the diversity of design alternatives by taking into account architectural features.

We envision three directions for future work.
First, we could extend the experimental design by incorporating additional interaction modes \cite{Ramirez_Romero_Simons_2019} (e.g., removing any of the objectives, or restricting the possible refactoring actions to apply) and assessing their effects on the process.
Secondly, we could introduce indicators to assess the quality of architectural models (e.g., like the spread indicator reported in~\cite{CortellessaDPT:ecsa2023}), thus enabling techniques that can help designers to get better insights into the characteristics of those models, and to make the search process more effective.
Lastly, another potential avenue would be to explore the weighted hypervolume approach, as discussed by Auger et al.~\cite{AugerBBZ09a}, where user preferences can be integrated by defining weight distribution functions. This would serve as an alternative way to allow the optimization process to prioritize certain regions of the Pareto front based on designer-defined preferences by quantitatively adjusting the importance of each objective.

\begin{acks}
\noindent Daniele Di Pompeo and Michele Tucci were supported by \SoBigDataITAck
Sebastian Frank was supported by \dqualizer.
The work of Pooyan Jamshidi has been supported in part by the National Science Foundation (Awards 2007202, 2107463, 2038080, and 2233873). J. Andres Diaz-Pace was supported by the project PICT-2021-00757, Argentina.
\end{acks}

\bibliographystyle{ACM-Reference-Format}
\bibliography{abbrev,biblio,references}

\end{document}